%% file: main.tex
\documentclass[preprint]{aastex63}

\newcommand{\sgr}{\mbox{SGR\,J1935+2154~}}
\newcommand{\sgrnos}{\mbox{SGR\,J1935+2154}}

\newcommand{\segone}{\mbox{Segment\,1~}}
\newcommand{\segtwo}{\mbox{Segment\,2~}}
\mathchardef\mhyphen="2D

\usepackage{amsmath}
\usepackage{amssymb}
\usepackage{graphicx}
\usepackage{color}

\shorttitle{\sgr Burst Forest}
\shortauthors{Kaneko et al.}

\graphicspath{{./}{figures/}}

\begin{document}

\title{ Fermi/GBM Observations of the \sgr Burst Forest}

\correspondingauthor{Y. Kaneko}
\email{yuki@sabanciuniv.edu}

\author[0000-0002-1861-5703]{Yuki Kaneko}\affiliation{Sabanc\i~University, Faculty of Engineering and Natural Sciences, \.Istanbul 34956 Turkey}

\author[0000-0002-5274-6790]{Ersin G\"o\u{g}\"u\c{s}}
\affiliation{Sabanc\i~University, Faculty of Engineering and Natural Sciences, \.Istanbul 34956 Turkey}

\author[0000-0003-4433-1365]{Matthew G. Baring} \affiliation{Department of Physics and Astronomy - MS 108, Rice University, 6100 Main Street, Houston, Texas 77251-1892, USA}

\author[0000-0003-1443-593X]{Chryssa Kouveliotou} \affiliation{Department of Physics, The George Washington University, 725 21st Street NW, Washington, DC 20052, USA}\affiliation{Astronomy, Physics, and Statistics Institute of Sciences (APSIS), The George Washington University, Washington, DC 20052, USA}

\author[0000-0002-0633-5325]{Lin Lin} \affiliation{Department of Astronomy, Beijing Normal University, Beijing 100875, China}

\author[0000-0002-7150-9061]{Oliver J. Roberts} \affiliation{Science and Technology Institute, Universities Space and Research Association, 320 Sparkman Drive, Huntsville, AL 35805, USA.}

\author[0000-0001-9149-6707]{Alexander J. van der Horst}\affiliation{Department of Physics, The George Washington University, 725 21st Street NW, Washington, DC 20052, USA}\affiliation{Astronomy, Physics, and Statistics Institute of Sciences (APSIS), The George Washington University, Washington, DC 20052, USA}

\author[0000-0002-7991-028X]{George Younes} \affiliation{Department of Physics, The George Washington University, 725 21st Street NW, Washington, DC 20052, USA}\affiliation{Astronomy, Physics, and Statistics Institute of Sciences (APSIS), The George Washington University, Washington, DC 20052, USA}

\author[0000-0001-9711-4343]{\"Ozge Keskin}
\affiliation{Sabanc\i~University, Faculty of Engineering and Natural Sciences, \.Istanbul 34956 Turkey}

\author[0000-0002-5887-6676]{\"Omer Faruk \c{C}oban}
\affiliation{Sabanc\i~University, Faculty of Engineering and Natural Sciences, \.Istanbul 34956 Turkey}

\begin{abstract} 
During 2020 April and May, \sgr emitted hundreds of short bursts and became one of the most prolific transient magnetars. At the onset of the active bursting period, a 130 s burst ``forest," which included some bursts with peculiar time profiles, were observed with the Fermi/Gamma-ray Burst Monitor (GBM). In this Letter, we present the results of time-resolved spectral analysis of this burst ``forest" episode, which occurred on 2020 April 27. We identify thermal spectral components prevalent during the entire 130 s episode; high-energy maxima appear during the photon flux peaks, which are modulated by the spin period of the source.  Moreover, the evolution of the $\nu F_{\nu}$ spectral hardness (represented by $E_{\rm peak}$ or blackbody temperature) within the lightcurve peaks is anti-correlated with the pulse phases extrapolated from the pulsation observed within the persistent soft X-ray emission of the source six hours later.  Throughout the episode, the emitting area of the high-energy (hotter) component is 1-2 orders of magnitude smaller than that for the low-energy component.  We interpret this with a geometrical viewing angle scenario, inferring that the high-energy component likely originates from a low-altitude hotspot located within closed toroidal magnetic field lines.

\end{abstract}


\section{Introduction} \label{sec:intro}

Among the populations of compact astrophysical objects, strongly magnetized isolated neutron stars, or magnetars \citep{DT92}, are by far the most vibrant. Their extreme magnetic fields (10$^{14}$-10$^{15}$ G) are likely powering the underlying mechanisms of the large variety of their observed characteristics \citep{TD95,TD96}. Since the discovery of magnetar-like field strengths with the Rossi X-ray Timing Explorer \citep{CK1998}, long-term monitoring of their bright, persistent, soft X-ray emission revealed new characteristics, such as excessive timing noise \citep{Woods2000} and timing anomalies \citep[glitches,][ and anti-glitches, \citealt{ach13,younes20d}]{kaspi00}. However, their persistent emission is not limited to only soft X-rays. Roughly a dozen magnetars have exhibited emission in hard X-rays (above 15 keV), and a handful of them have been observed in optical and/or radio wavelengths. See \cite{kasbel17} for a recent review covering their broadband properties, and \cite{olausen} for a dynamic web portal listing the properties of currently known magnetars.

What remains the most intriguing property of some magnetars is the repeated emission of multiple, highly energetic bursts. Magnetar bursts appear in broad morphological and energetic varieties; these range from sub-second-long typical short bursts at super-Eddington luminosities (10$^{39}$--10$^{41}$\,erg s$^{-1}$) to giant flares that last up to several minutes, with energies in excess of 10$^{45}$\,erg. In between these classes is a group of bursts, each typically with a duration of a few to a few tens of seconds and energies of around 10$^{42}$\,erg, which are commonly referred to as intermediate events \citep{Kouveliotou2001}. Such events have been seen from SGR 1627$-$41 \citep{mazets99}, SGR 1900+14 \citep{ibrahim,lenters,israel08}, SGR 1806$-$20 \citep{gogus11}, and SGR J1550$-$5418 \citep{mereg09,kaneko10}. These events usually occur at the onset, or in the early phases, of their burst active episodes. Spectrally, the intermediate events exhibit thermal characteristics, represented by a blackbody temperature of a few keV, which decays over a timescale of seconds to minutes. Pulsations are commonly observed during the tail of the decaying emission, with a higher pulsed intensity compared to the level prior to the intermediate event. The case of SGR J1550$-$5418 was particularly interesting, as that burst ``storm" occurred several hours before the intermediate events \citep{mereg09}. At the onset of this storm, an enhancement of the underlying persistent emission in hard X-rays lasted $\sim2$ minutes. The spectrum of this enhanced emission was well described with a blackbody temperature of $kT=17$\,keV, accompanied with pulsed emission detected up to $\sim$70~keV \citep{kaneko10}. The enhanced emission was energetically comparable to that of emission from tails of typical intermediate events, and was interpreted as emission originating from a short-lived hot spot on the surface of the neutron star.

\sgr became active again in 2020 April when it emitted hundreds of bursts \citep{lin20}. The onset of this episode was marked with exceptional activity; a storm of highly energetic bursts detected with the Swift/Burst Alert Telescope  \citep[BAT:][]{palmer20} and the Fermi/Gamma-ray Burst Monitor \citep[GBM;][]{fletcher}. Follow-up observations of \sgr with Neutron star Interior Composition ExploreR (NICER) captured another cluster of bursts $\sim$6 hours later \citep{younn20}, while INTEGRAL, KONUS-WIND and HXMT captured the X-ray counterpart of the first observed Galactic Fast Radio Burst \citep{mereghetti20,ridnaia2021,Li2021}. A careful inspection of the Fermi/GBM data showed the onset of the event resembled the 2006 ``burst forest" observed from SGR\,$1900+14$, during which more than 40 bursts were detected with Swift/BAT over $\sim30$\,s \citep{israel08}. Nearly one-third of those bursts exhibited highly unusual morphology with durations of $\sim$1\,s, longer than typical magnetar bursts, some of which were flat-topped \citep{israel08}. We also note that a similar burst forest from SGR\,1806$-$20 was observed with INTEGRAL, during which $>$100 bursts were observed within $\sim$600\,s \citep{gotz06}.

\citet{lin20} analyzed time-integrated spectra of 125 Fermi/GBM bursts from the 2020 April activity of \sgr, excluding the 130\,s segment, where a cluster of such peculiar events were observed, hereafter referred to as the ``burst forest." Here we present the results of our time-resolved spectral analyses of this 130\,s long burst forest, observed with Fermi/GBM. The observation and the data analysis methods are described in \S{\ref{sec:data}}, followed by the presentation of the analysis results in \S{\ref{sec:res}}.  We interpret and discuss the results in \S{\ref{sec:disc}}. We note that the estimated distance to \sgr ranges from $\sim$2 to 9\,kpc \citep[see][and the references therein]{zhong20} including the most recent estimations from the dust-scattering halo of 4.5$^{+2.8}_{-1.3}$\,kpc \citep{mereghetti20} and from the host supernova remnant of 6.6$\pm$0.7\,kpc \citep{zhou20}; here we use the distance of 9\,kpc, consistent with \citet{lin20}.

\section{Observation and Data Analysis} \label{sec:data}
\begin{figure*}
\begin{center}
\includegraphics[scale=1, trim=45 220 30 320, clip]{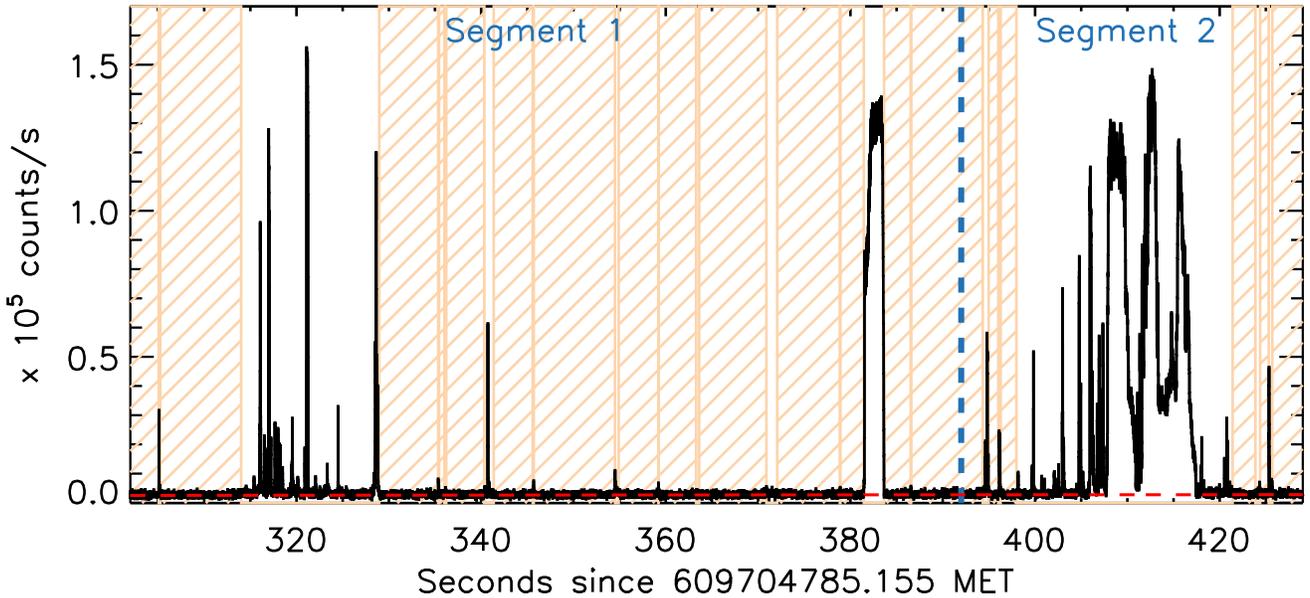}
\end{center}
\caption{
Lightcurve of the entire ``burst forest" episode ($T_0 + 310$\,s to $T_0 + 430$\,s, 8 ms resolution). The count rates are summed over all triggered detectors. The blue dashed line shows the separation between Segments 1 and 2. The yellow-hatched regions are the background time intervals used to determine the background rate (red dashed line).  
\label{fig:lc_2segs}}
\end{figure*}

GBM consists of 12 NaI detectors (covering 8\,keV-1\,MeV) and 2 BGO detectors ($\sim$200\,keV-30\,MeV). On 2020 April 27, three of the NaI detectors (\#6, 7, \& 9) were triggered at 18:26:20 UT (609704785.155 MET = $T_0$) by a burst originating from the direction of \sgr (trigger\# bn200427768).  Upon inspection of the event data, we identified the burst forest from $T_0 + 310$\,s to $T_0 + 430$\,s (see Figure~\ref{fig:lc_2segs}), which included several broad peaks with a few-seconds duration. The GBM triggered data are accumulated in several different formats.  Of those, the time-tagged event (TTE) datatype provides the finest resolution, both in time  (2\,$\mu$s) and in energy (128 energy channels).  Therefore, we used the TTE data with the time-dependent detector response matrices (rsp2) provided by the GBM team. For this work, we did not include the BGO data as adding the BGO data did not affect the spectral parameters at all due to the low count rates above a few hundred keV.

\subsection{Detector selection and background modeling}
Because the detector-to-source angles change considerably during 130\,s episode, we divided this data into two segments separated at $T_0 +392$\,s, taking into account the detector angles and the source lightcurve. We also calculated the field of view and spacecraft blockage (mainly by the Large Area Telescope) for each NaI detector during this time period. We used detectors having detector-to-source angles $< 55^{\rm o}$ that were unblocked for each of the two segments (1.~$T_0+310$\,s to $T_0+392$\,s and 2.~$T_0+392$\,s to $T_0+433$\,s). These requirements constrained the detectors used for the analysis to NaI\,6, 7, and 9 for \segone and NaI\,0, 1, 6, and 9 for Segment\,2.

To model the background count rates, we first generated a Bayesian-block representation of the lightcurve \citep{scargle13} using the TTE data of the triggered detectors (binned to 8\,ms, 10--100\,keV) and defined the background time intervals as the blocks longer than 4\,s (hatched regions in Figure\,\ref{fig:lc_2segs}). The background count rates were then interpolated using the 8 ms lightcurve of the background time intervals in the energy range used in the spectral analysis (8--500\,keV), by fitting a second-order polynomial function. We also cross checked our choice of the background time intervals by running the untriggered event search, which uses Poisson statistics to identify count rate increases, over the entire burst-forest period; none of the untriggered bursts identified were within our background time intervals.

\subsection{Evaluation of Deadtime and Pile-up}
Before proceeding with the spectral analysis, we evaluated the possibility of count saturation in the data due to the very high count rates recorded for some of the burst forest peaks. The TTE data suffers from a deadtime of 2.6\,$\mu$s due to the fixed data-packet processing speed of 375~kHz on the spacecraft~\citep{meegan09}. Consequently, bandwidth saturation of the data occurs when the collective count rates of all 14 GBM detectors (12 NaI + 2 BGO), exceed this limit. For the burst forest episode, the maximum count rate summed for all 14 detectors over all energy bands is $\sim250$\, kHz, which is well below this saturation limit.
Additionally, we also evaluated the data for pulse pile-up effects, a phenomenon that inaccurately processes normally discrete spectral and temporal values when the ``recovered" count rates exceed $\sim$10$^5$ counts per second per detector \citep{chaplin13, bhat14}. The deadtime-corrected TTE count rates for the detectors used in this study do not exceed this limit, and thus we conclude that our data is unaffected by this behavior and our analysis of the burst forest is accurate.

\subsection{Spectral Analysis}
We performed time-resolved spectral analysis  ($8-$500\,keV) using the RMFIT spectral analysis tool (version 4.3.2).\footnote{https://fermi.gsfc.nasa.gov/ssc/data/analysis/rmfit} The energy range of 30-40\,keV was excluded in our analysis to avoid possible contribution to the fit statistics due to the iodine $K$-edge at 33\,keV. This exclusion, however, did not affect the analysis outcome.
We fit each of the two time segments with four models commonly used to describe spectra of magnetar bursts: a power law with exponential cutoff (COMPT), a single blackbody (BB), a double blackbody (BB+BB), and a blackbody plus power law (BB+PL). The fit parameters were obtained by minimizing the log-likelihood (C-stat).  The spectral fits were performed in various time resolutions: 4, 16, and 128 ms.  The 4 ms and 16 ms spectral bins were created with a signal-to-noise (S/N) minimum of 25 in the brightest detector. We present in Table\,\ref{tab:param} the spectral parameters obtained with either COMPT, BB, or BB+BB fits for each time interval. We note here that based on the spectral fit results, we identified two distinct kinds of intervals, which we define as peaks and troughs. The former are invariably associated with the higher photon fluxes and can be best fit with either COMPT or BB+BB. The latter are associated with flux minima and are best fit with either COMPT or a single BB. Because the COMPT model provides reasonable fits for all spectra of the burst forest, we use the BB+BB and BB fits to define the peak/trough definition. 

Throughout this Letter we have consistently used equal time bins of 128\,ms for the peaks in all the figures, unless otherwise stated. However, for troughs, the requirement of a 25 S/N results in bins with variable time resolution. All these parameters are presented in Table\,\ref{tab:param}. Finally, we find that the fluence during the entire duration of the burst forest was $F = (2.305 \pm 0.001)\times 10^{-4}$\,erg\,cm$^{-2}$ in the $8–500$ keV energy range; Segment 1 accounted for $(0.562 \pm 0.002) \times 10^{-4}$\,erg\,cm$^{-2}$ and Segment 2 for $(1.743\pm 0.002) \times 10^{-4}$ \,erg\,cm$^{-2}$. 

\section{Results}\label{sec:res}
\begin{figure*}
\begin{center}
\epsscale{1.16}
\plottwo{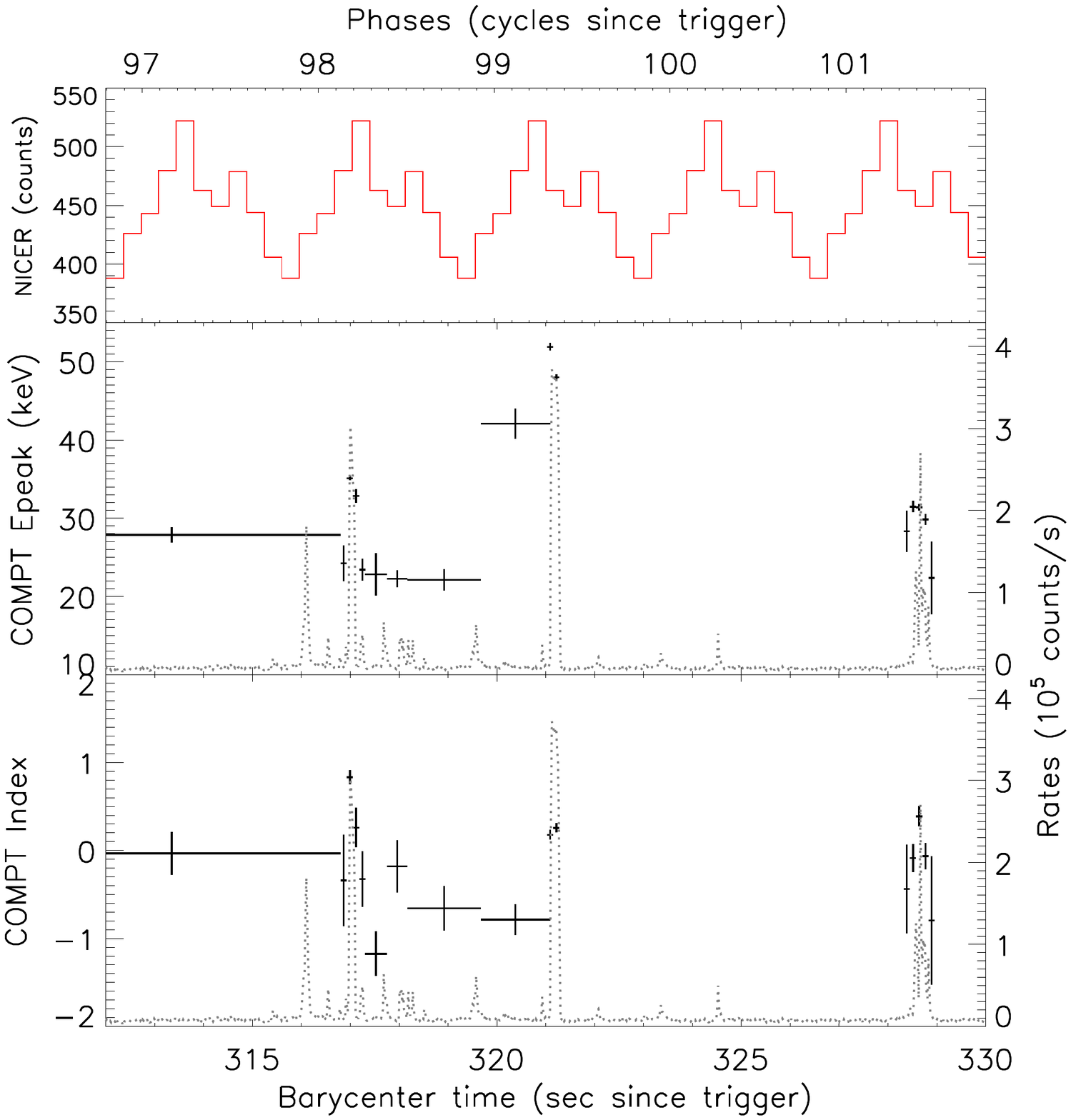}{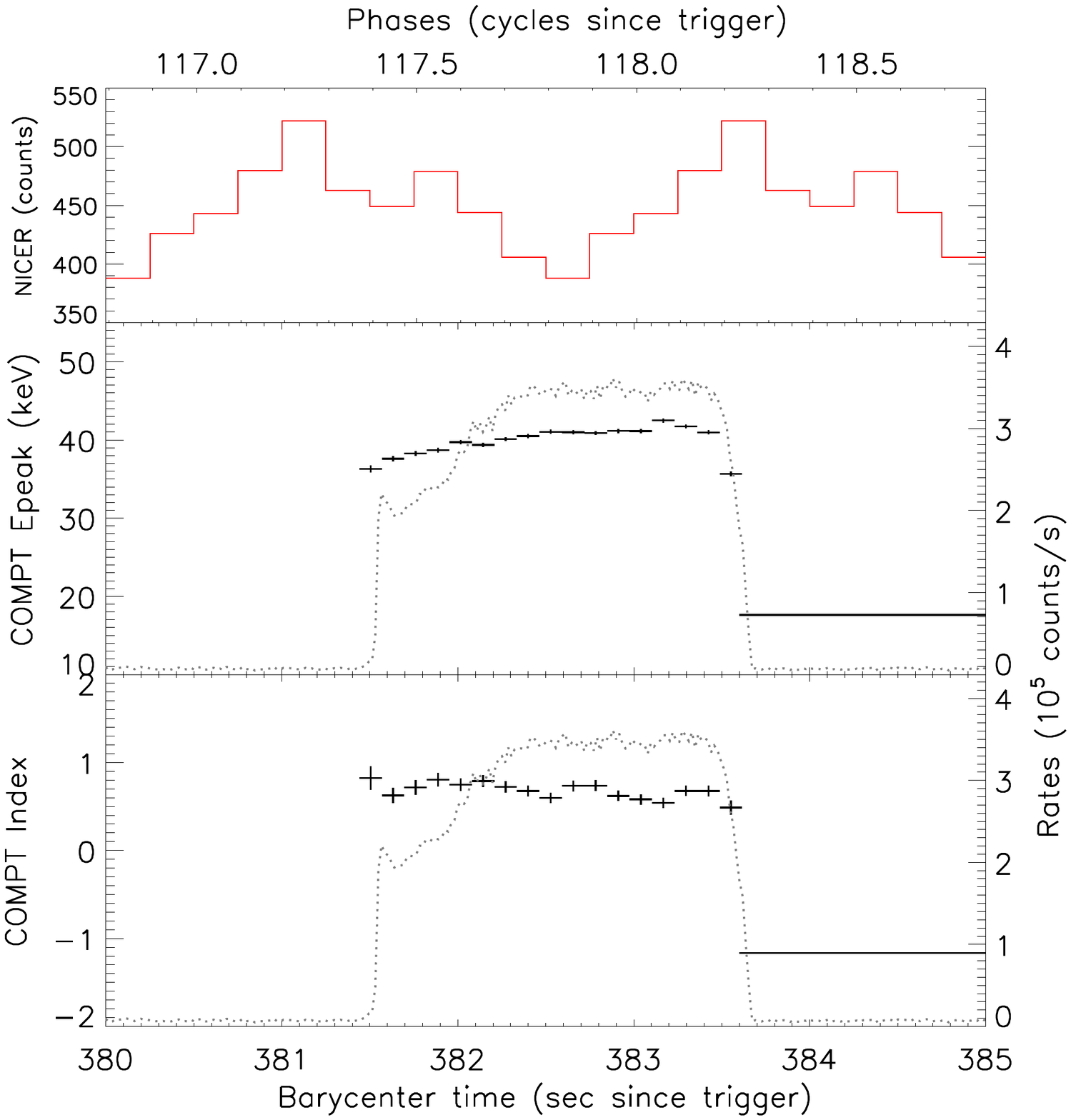}
\includegraphics[scale=0.5, trim=0 150 0 50, clip]{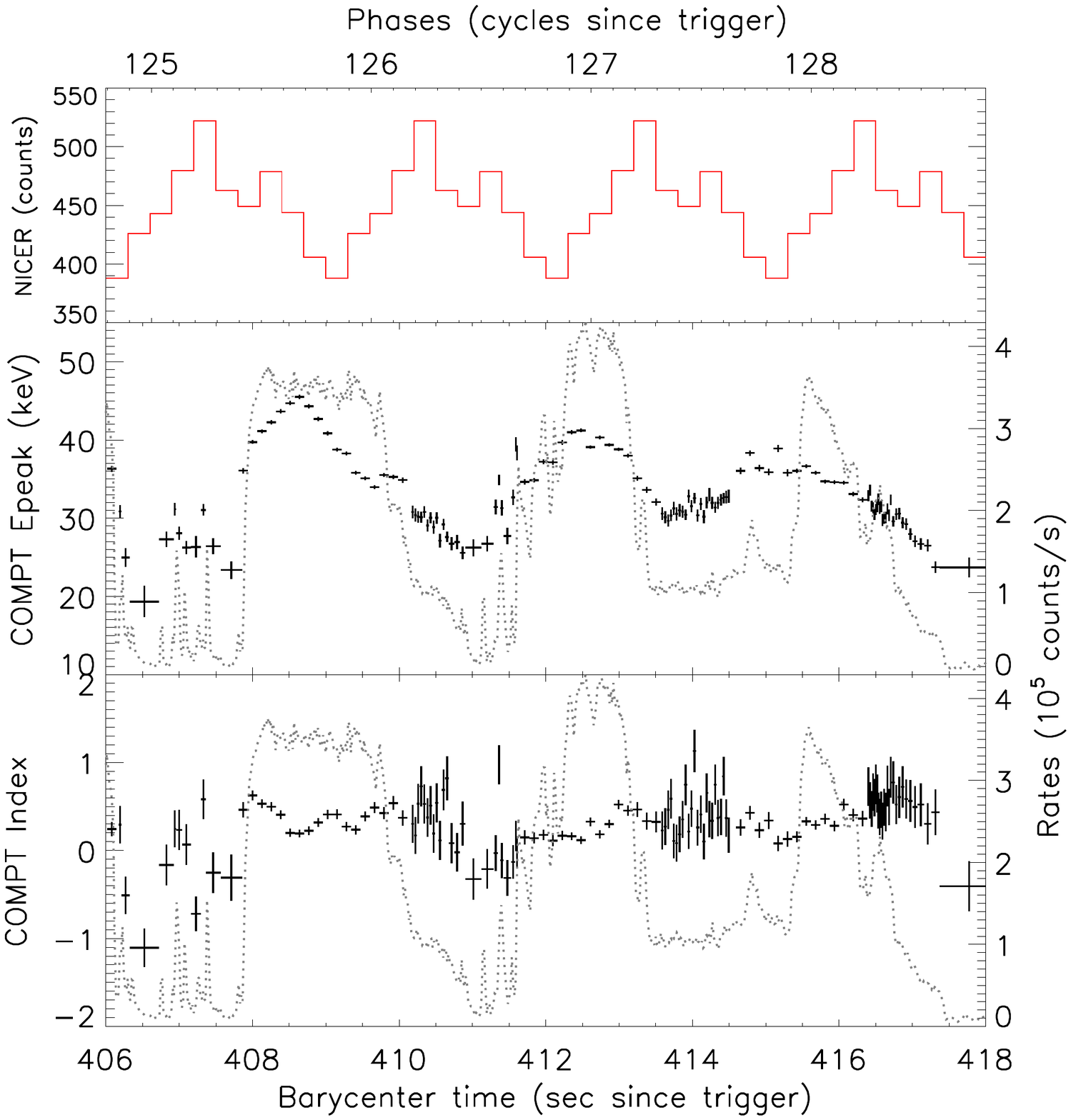}
\end{center}
\caption{COMPT-fit parameters ($E_{\rm peak}$ and index $\Gamma$) for \segone (top two panels) and for the triple peaks of \segtwo (bottom panel; 4 ms resolution binned to S/N $\geq$ 25 for troughs and 128 ms resolution for peaks).  The count rates (right y-axis) are plotted with gray dotted lines. The red histograms at the top of each panel show the extrapolated NICER phase cycles; the cycle numbering starts at the GBM trigger time (see \S\ref{sec:disc}).
\label{fig:compt_param}}
\end{figure*}

We find that of the four models mentioned in Section 2.3, the parameters for three of these (COMPT, BB, and BB+PL) are reasonably constrained for {\it all} time-resolved spectra.  When we fit the BB+BB model to the troughs, it often fails to constrain the high-energy BB parameters. Finally, the BB+PL model provides the least acceptable statistics and has not been used in further analysis (see also Section\,\ref{sec:bic}). When fitting the COMPT model, we find that the $\nu F_{\nu}$ spectral peak ($E_{\rm peak}$) varies significantly within each individual temporal peak, while the low-energy spectral index remains constant around 0.5 (see Figure\,\ref{fig:compt_param}). The $E_{\rm peak}$ generally follows the lightcurve, and we find a significant positive correlation between $E_{\rm peak}$ and photon flux; the Spearman rank-order correlation coefficients are $r_s = 0.4$ and 0.9, with $P < 10^{-6}$ and $10^{-17}$ for the \segone and Segment 2, respectively.  We note that the COMPT model is defined as $f(E) \propto (E/E_{\rm peak})^{\Gamma}$; therefore, $\Gamma > 0$ indicates that the spectral curvature below $E_{\rm peak}$ resembles that of thermal spectra.  The thermal nature of these spectra is also supported by the fact that BB+PL fits provide well-constrained parameters: the PL index remains constant (weighted average of $-2.35 \pm$0.07 for both segments combined), representing the spectra above $\sim$100\,keV, and the BB $kT$ follows the $E_{\rm peak}$ evolution.

\begin{figure*}
\begin{center}
\epsscale{1.16}
\plottwo{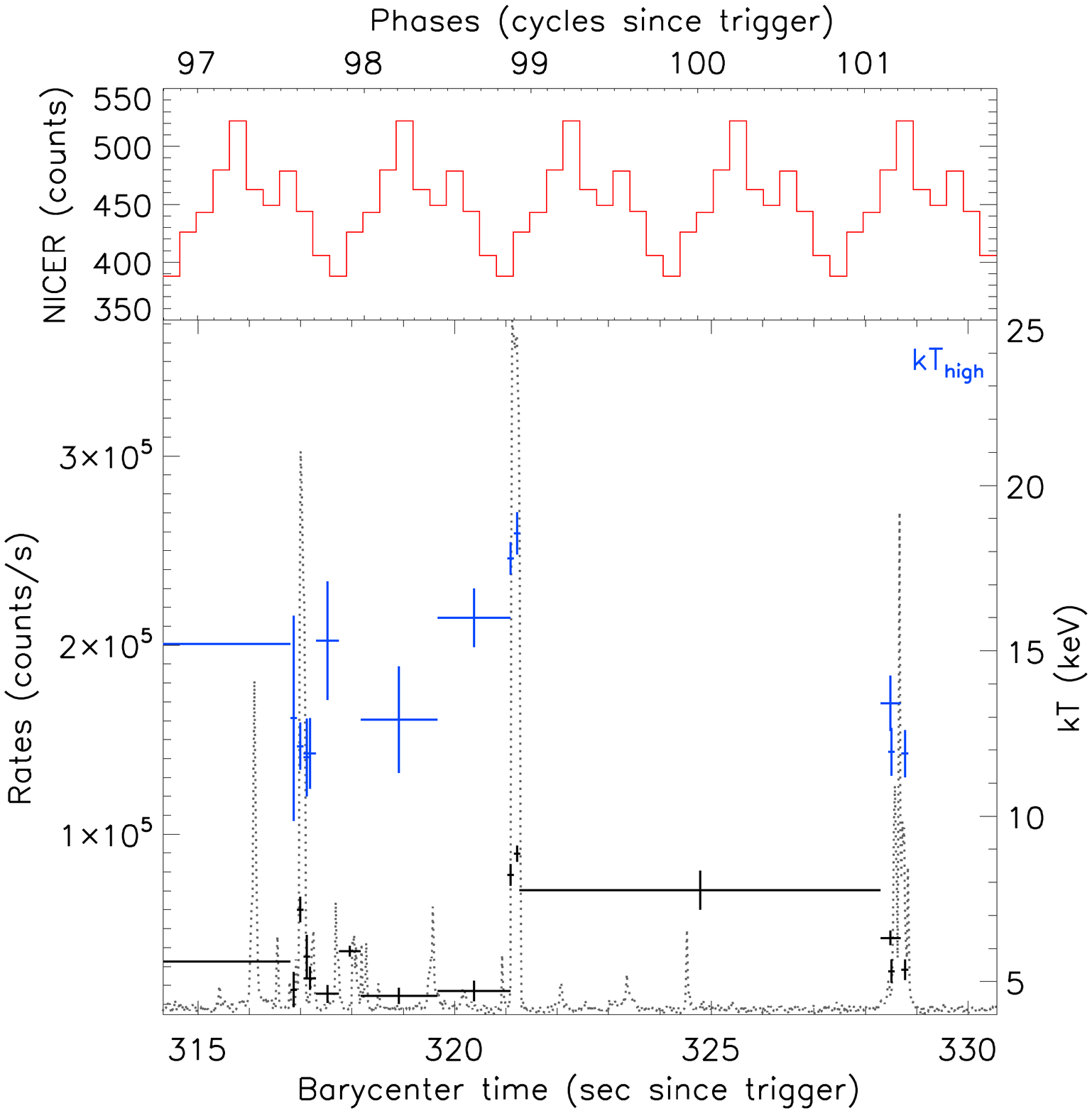}{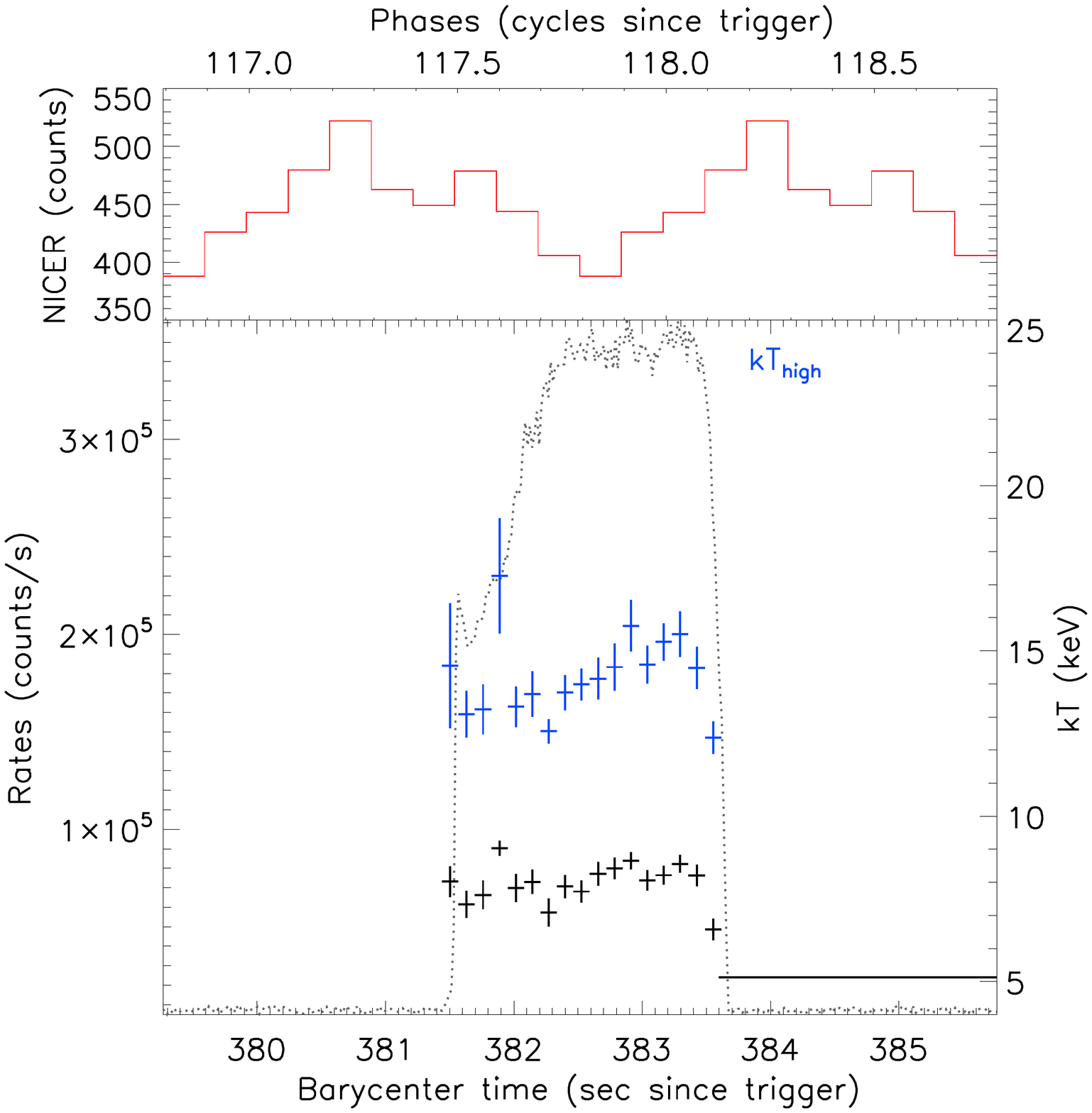}
\end{center}
\vspace{-2cm}
\caption{Blackbody $kT$ evolution during Segment 1 (4 ms resolution binned to S/N $\geq$ 25 for troughs and 128 ms resolution for peaks). $kT_{\rm high}$ is shown in blue for the intervals where BB+BB better represents the spectra over a single BB.  $kT_{\rm low}$ and single BB $kT$ are shown in black. The count rates (left y-axis) and the temperatures (right y-axis) are shown vs. barycentered times along with the phases of the persistent emission pulse profiles extrapolated using NICER data (top panels).
\label{fig:bbbb_kt_1}}
\end{figure*}

When we fit the BB+BB model, we find that within the Segment 1 peaks and troughs, the low and high $kT$ do not vary drastically; however, for Segment 2 peaks the $kT$s vary significantly. The $kT$s are also strongly correlated with each other in all peaks in Segment\,2, with $r_s \sim 0.8$ and $P < 10^{-23}$.  In Figures\,\ref{fig:bbbb_kt_1} and \ref{fig:bbbb_kt_2}, we present the $kT$ evolution during the two segments.

\subsection{Spectral Fit Statistics}\label{sec:bic}

\begin{figure*}
\begin{center}
\includegraphics[scale=0.6, trim=0 220 0 0, clip]{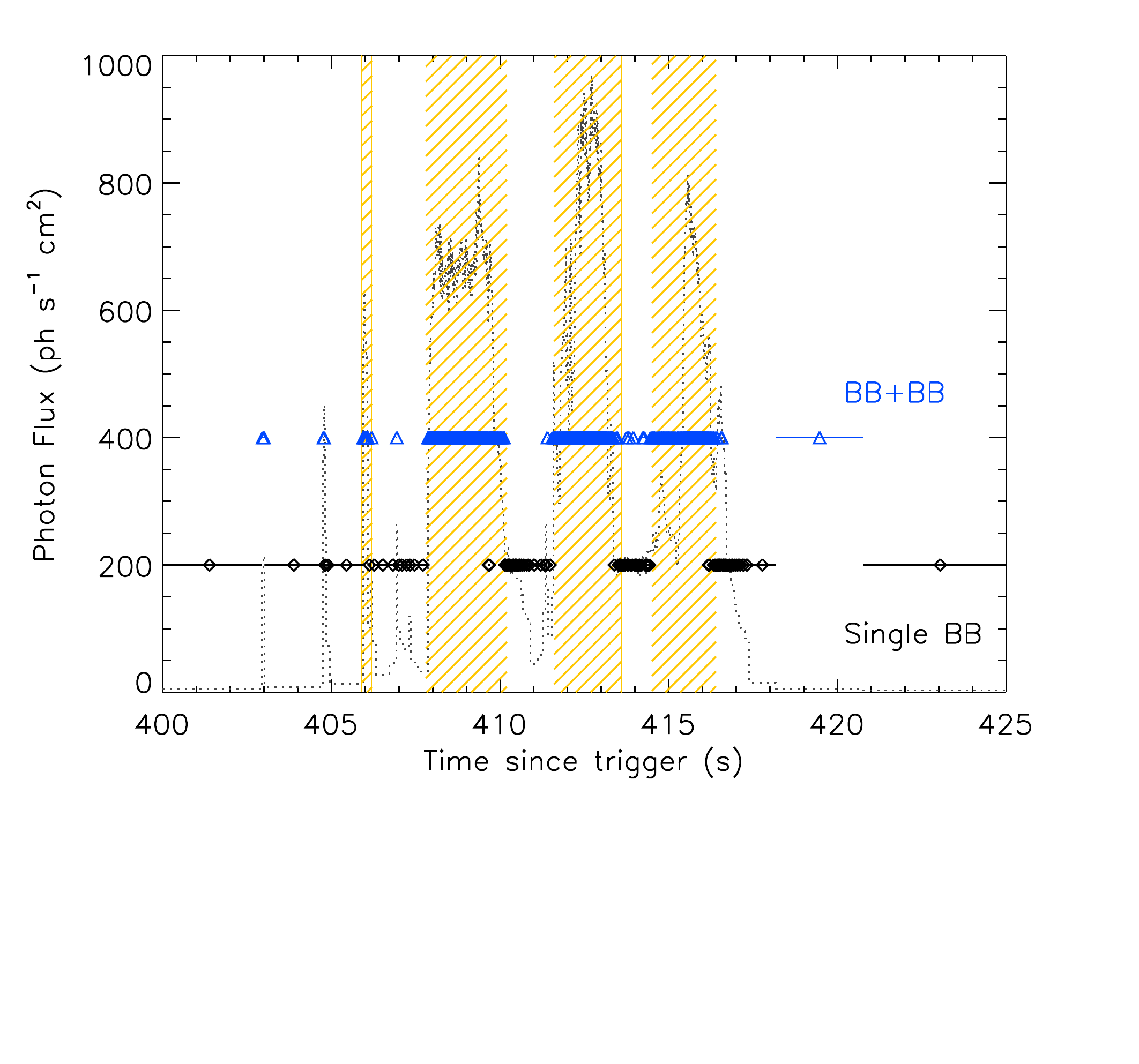}
\end{center}
\vspace{-3.5cm}
\caption{Segment\,2 Photon flux history showing the preferred model between BB and BB+BB for each of the time intervals (4-ms resolution binned to S/N $\geq$ 25) determined based on the differences in BIC (i.e., $\Delta {\rm BIC = BIC_{BB} - BIC_{BB+BB}} >$ 10 if BB+BB is preferred).  The y-axis values for BB+BB or single BB points were arbitrarily selected. The hatched regions define the ``peaks" in our analysis, as discussed above. The photon flux values are estimated with the fitted models.
\label{fig:bic_comparison}}
\end{figure*}

The fit statistics that we employed for our spectral fitting (C-stat) does not provide the goodness of fit and cannot be directly used to identify the model that better describes each of the spectra.  Therefore, we calculate for each spectral fit the Bayesian Information Criterion  \citep[BIC;][]{liddle07} as follows:

\begin{equation}
  {\rm BIC} = -2 \ln \mathcal{L}_{\rm max} + k \ln N = {\rm C\mhyphen stat} + k \ln N , 
\end{equation}

\noindent where $\mathcal{L}_{\rm max}$ is the maximum likelihood, $k$ is the number of parameters in the photon model and $N$ is the number of data points.  We then compare the BIC values associated with each of the four model fits in pairs and the model with significantly lower BIC, i.e. with $\Delta {\rm BIC} > 10$, was considered preferred.
According to our $\Delta {\rm BIC}$ comparisons, we find that BB+BB for the peaks and BB for most of the troughs are consistently identified as preferred models.  In Figure~\ref{fig:bic_comparison}, we show the preferred models for all spectra in  Segment\,2.  We see a clear transition of the preferred models from BB+BB to BB in the decaying parts of the broadest peaks.

\begin{figure*}
\begin{center}
\plotone{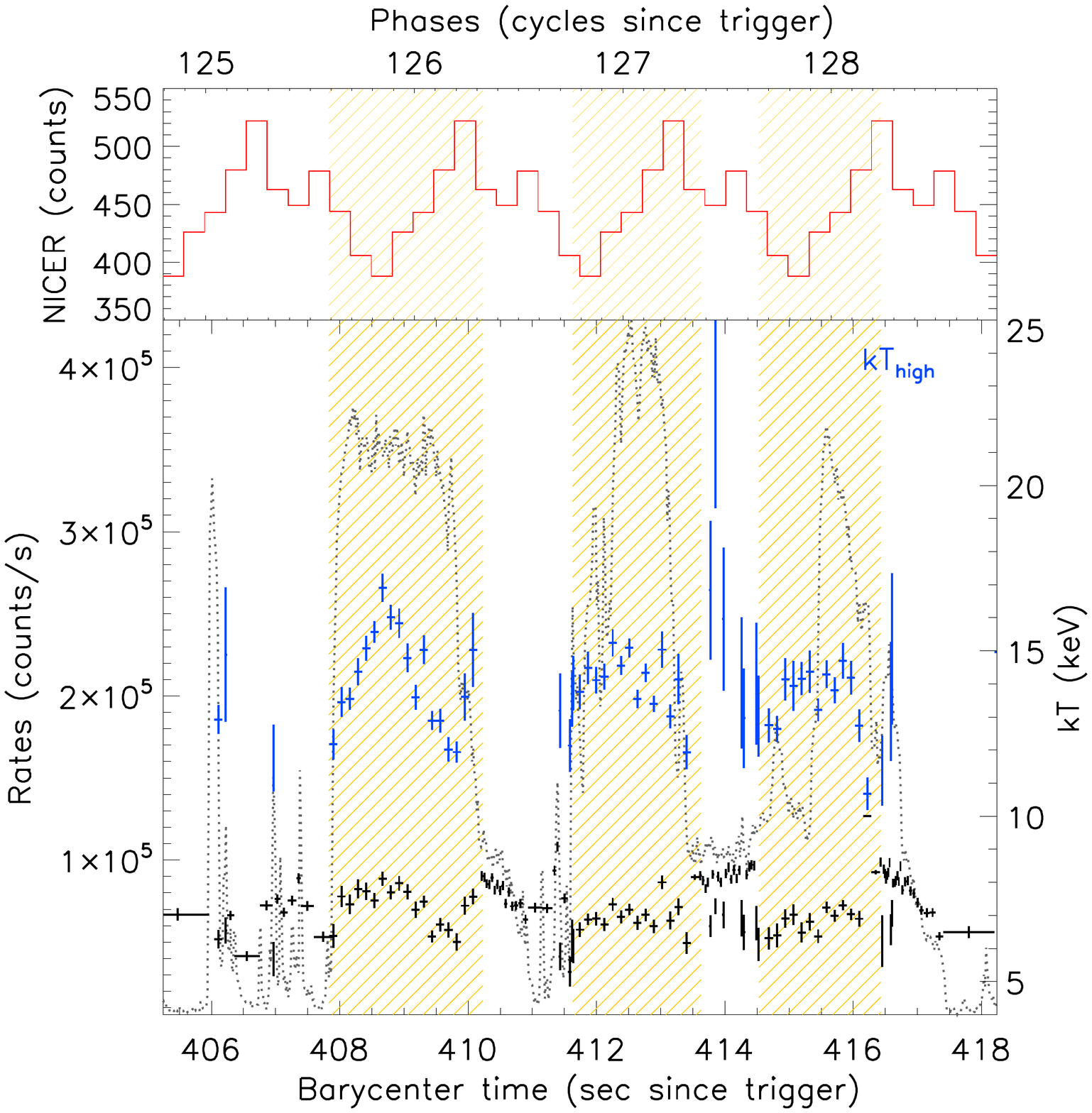}
\end{center}
\vspace{-4cm}
\caption{Blackbody $kT$ evolution during \segtwo (4 ms resolution binned to S/N $\geq$ 25 for troughs and 128 ms resolution for peaks). $kT_{\rm high}$ is shown in blue for the intervals where BB+BB better represents the spectra over a single BB.  $kT_{\rm low}$ and single BB $kT$ are shown in black.  The count rates (left y-axis) and the temperatures (right y-axis) are shown vs. barycentered times, along with the phases of the persistent emission pulse profiles extrapolated using NICER data (top panels). The yellow shaded intervals indicate the intervals where BB+BB is consistently preferred over BB (see Figure \ref{fig:bic_comparison} ).
\label{fig:bbbb_kt_2}}
\end{figure*}

Finally, within the trough spectra where BB is preferred, the $kT$ does not vary significantly. The weighted means of the BB $kT$ in the Segment\,1 and 2 trough spectra are 8.0$\pm$0.5\,keV and 7.8$\pm$0.3\,keV, respectively.  
We assume, therefore, that this component is manifested as the low-energy $kT$ in the BB+BB spectra, and fix the low-energy $kT$ to the weighted mean values mentioned above. We then
proceed to study the evolution of the  high-energy $kT$.  Although in some parts of the peaks the high-energy $kT$ is then found to be slightly lower ($\lesssim 2$\,keV) than those found with the varying low-energy $kT$, we find that the overall high-$kT$ evolution does not change significantly from what is seen in Figure\,\ref{fig:bbbb_kt_2}.

\section{Discussion}\label{sec:disc}

\begin{figure*}
\begin{center}
\includegraphics[scale=0.7]{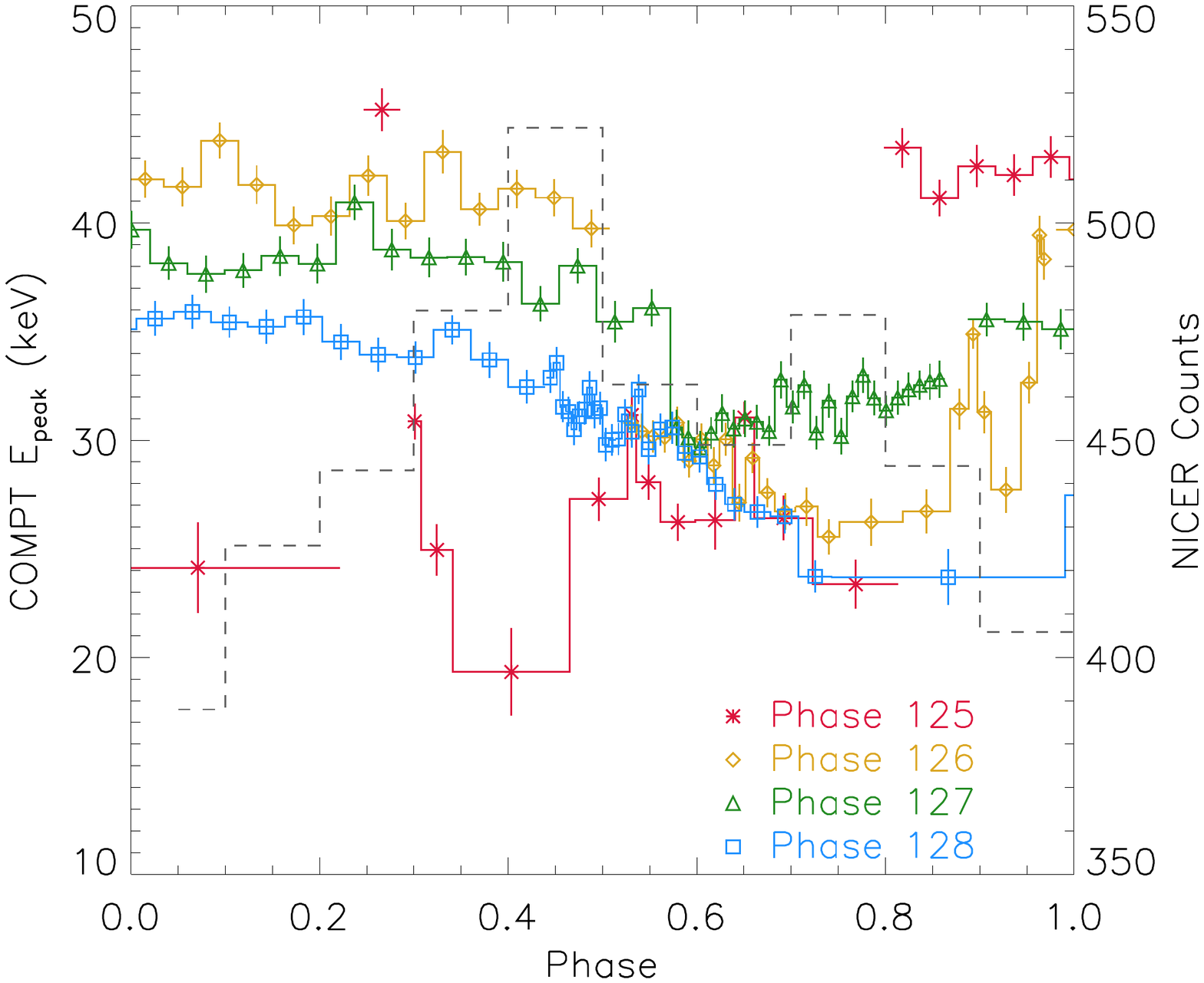}
\end{center}
\vspace{-6cm}
\caption{COMPT $E_{\rm peak}$ values of \segtwo peaks (Figure\,\ref{fig:compt_param}, bottom panel) as a function of phase. The NICER pulse phase is extrapolated to the Segment 2 data, shown with black dashed lines (right axis; NICER pulse count rate, also shown in Figure\,\ref{fig:bbbb_kt_2}) The colored histograms correspond to the evolution of $E_{\rm peak}$ in different phases.
\label{fig:compt_ep_phase}}
\end{figure*}

\citet{younes20} presented the pulse profile of the  persistent emission of \sgr as observed with NICER $\sim$6\,hours after the GBM burst forest episode discussed here.  We converted the GBM photon arrival times during the burst forest to barycentric times and matched the episode lightcurves and their spectral parameter evolution with the extrapolated NICER pulse profile.  The extrapolated pulse profile is shown in the top panels of Figures\,\ref{fig:compt_param}, \ref{fig:bbbb_kt_1}, \ref{fig:bbbb_kt_2}. In particular, we observe from the one-component COMPT model fits (Figure\,\ref{fig:compt_param}) that the $E_{\rm peak}$ values of the spectra appear to anti-correlate with the pulse profile, with the spectra peaking at higher energies at or close to the minima of the pulse profile.  Furthermore, we find that in Segment\,2, the transition from peaks to troughs occurs roughly at the spin period ($P_{\rm spin}=3.24$\,s) of the source (see hatched regions in Figure\,\ref{fig:bbbb_kt_2}). Motivated by this trend, we superimpose the COMPT $E_{\rm peak}$ evolution of each pulse phase within \segtwo to investigate the phase-resolved spectral evolution, which is shown in Figure\,\ref{fig:compt_ep_phase}.  It is evident that for all three peaks, the $E_{\rm peak}$ stays high ($\sim$45\,keV) during the phase minimum and transitions to lower energies ($\sim$25--30\,keV) at the phase maximum.  It is also clear from the figure that right before and right after the triple peaks in the lightcurve (i.e., beginning of Phase~125 and the end of Phase~128), the spectra do not follow this trend. We also note that there are some time intervals in the trough parts of the lightcurve (i.e., non-hatched regions) for which the spectra are still better described with the BB+BB model, based on the BIC comparisons.

\begin{figure*}
\begin{center}
\epsscale{1.16}
\plottwo{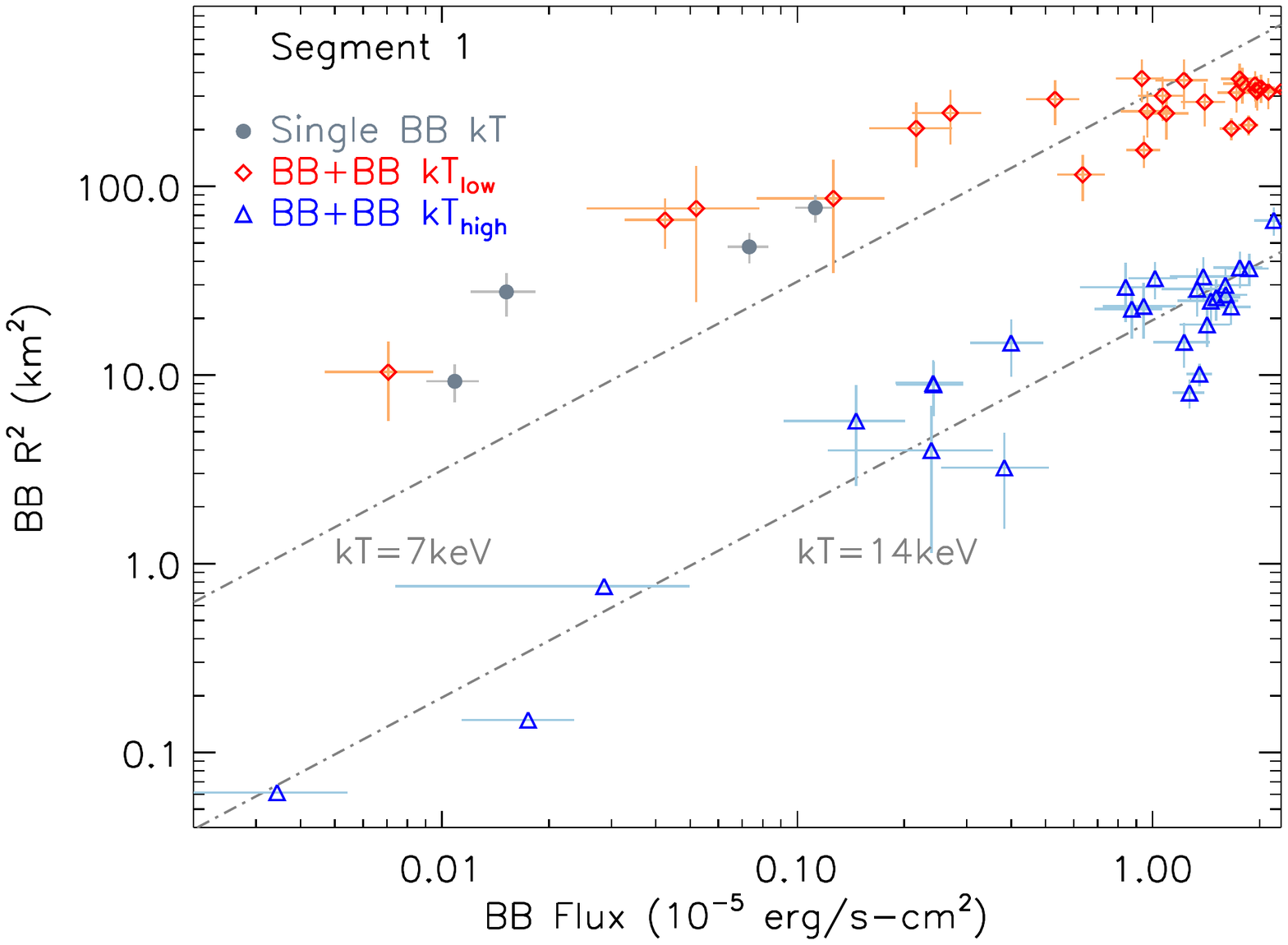}{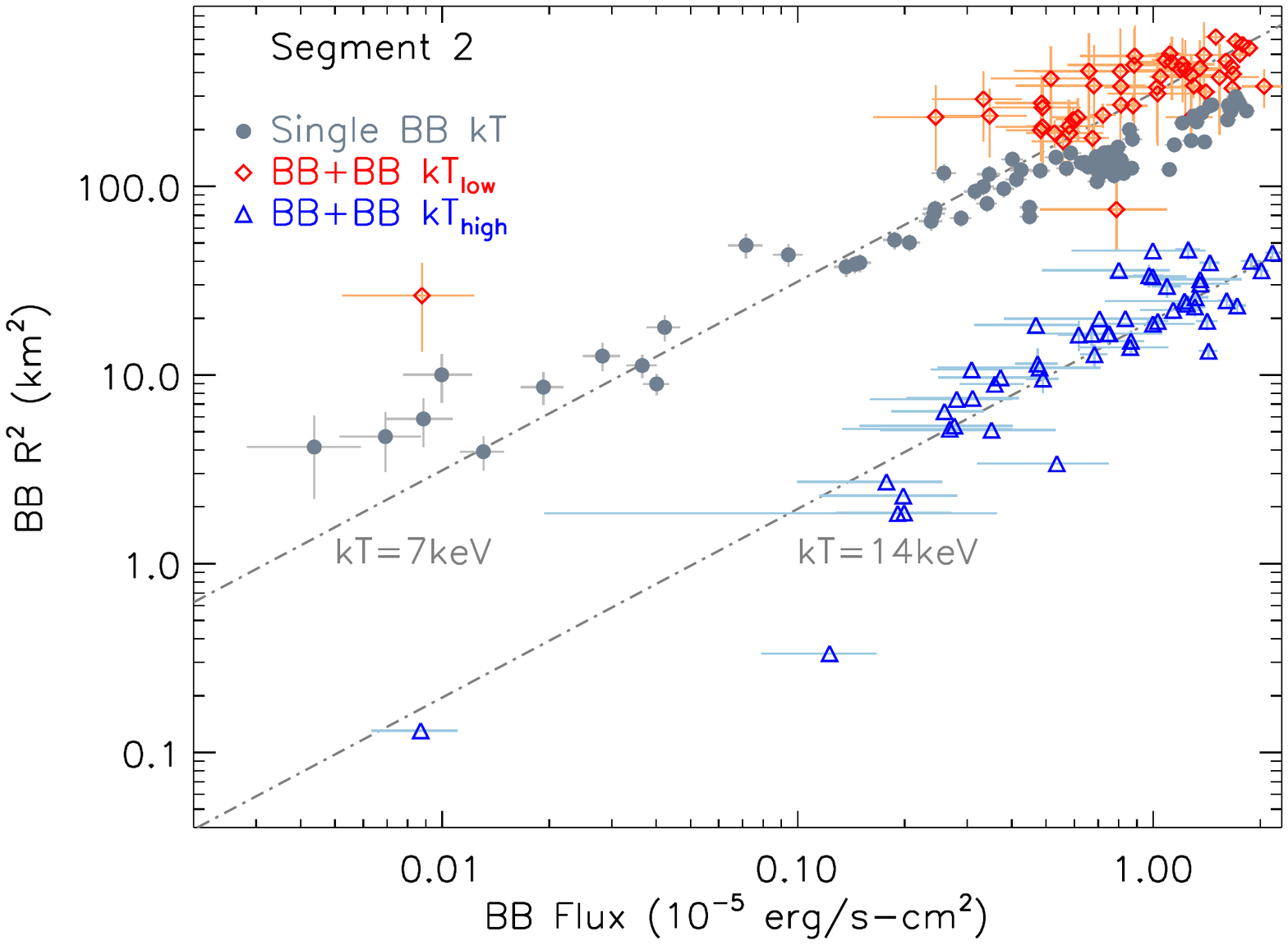}
\end{center}
\vspace{-3cm}
\caption{BB \& BB+BB flux vs.~emitting areas ($R^2$) for \segone and \segtwo separately. Lines corresponding to constant temperatures, $kT = 7$\,keV \& 14\,keV are shown as a reference in both panels.
\label{fig:flux_rad_segs}}
\end{figure*}

\begin{figure*}
\begin{center}
\includegraphics[scale=0.8, trim=0 100 0 100, clip]{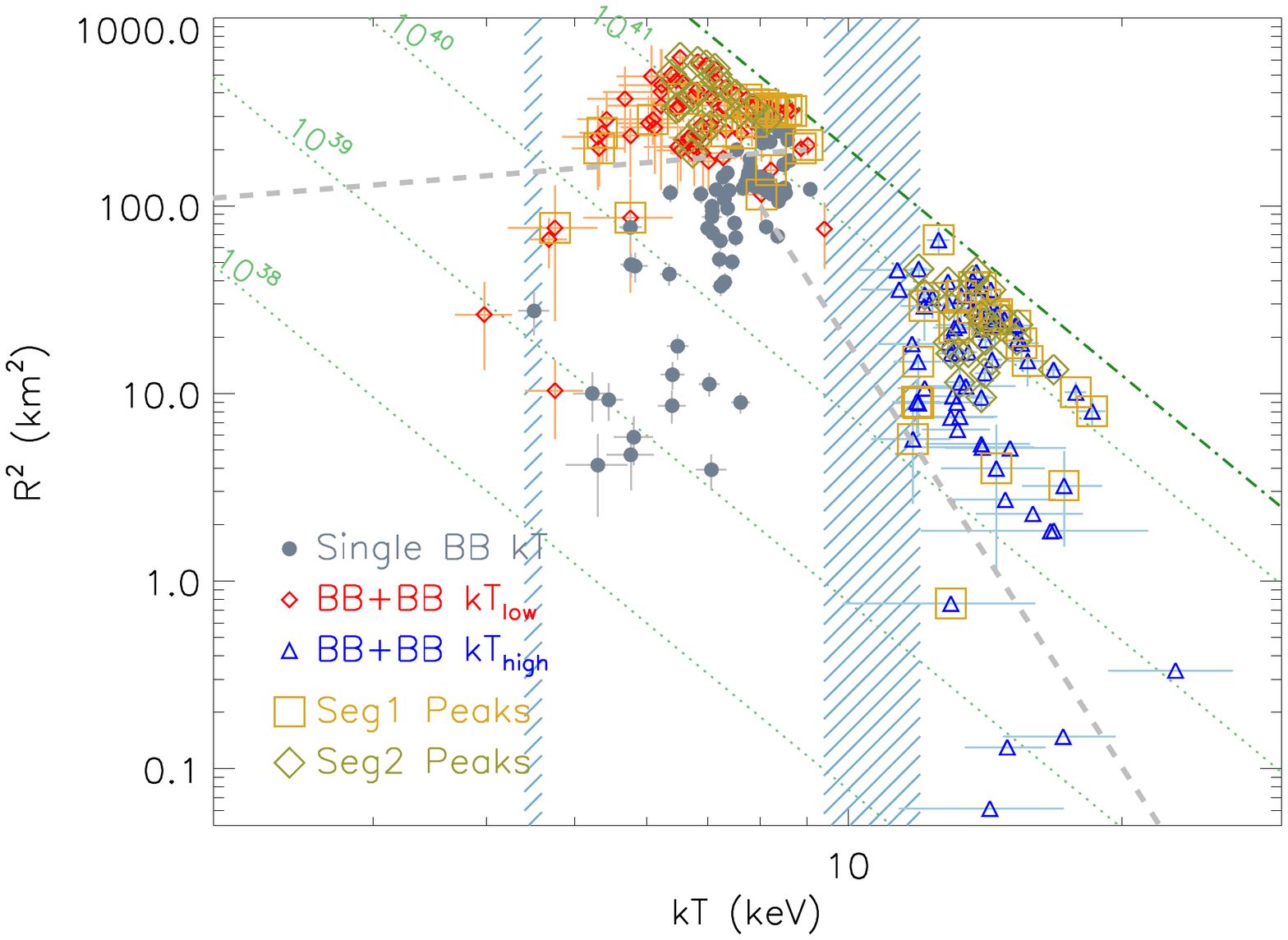}
\end{center}
\vspace{-4cm}
\caption{BB \& BB+BB emitting areas ($R^2$) vs.~$kT$. The $kT_{\rm Low}$ are shown in red, while $kT_{\rm High}$ are shown in blue.  Large squares and diamonds indicate peak parameters in Segments 1 and 2, respectively.  Gray dashed lines show the correlations reported in Figure\,3 of \citet{lin20} from the analysis of all bursts within the same activation period of the source; the blue shaded regions show the mean values of low and high BB kT in the \citet{lin20} burst sample. The green dashed-dotted line shows the $R^2 \propto L(kT)^{-4}$ line corresponding to a luminosity upper limit of $2.6 \times 10^{41}$\,erg\,s$^{-1}$.  The isoluminosity contours, corresponding to $L=10^{38}-10^{41}$\,erg\,s$^{-1}$, are shown as light green dotted lines.
\label{fig:r2_kt}}
\end{figure*}

We present in Figure\,\ref{fig:flux_rad_segs} the the energy flux versus the emitting radius of the BB+BB and BB components. We note that especially for Segment 2, the single BB $kT$ (i.e., the trough spectra) remains constant around 7\,keV, while the flux varies around two orders of magnitude.  The $kT_{\rm high}$ also remains roughly constant at around 14\,keV throughout both segments. The high-energy BB corresponds to smaller emitting radii than the low-energy component, while the ranges of the energy fluxes of the four peaks are comparable. In Figure\,\ref{fig:r2_kt}, we present the estimated BB emitting area ($R^2$) as a function of temperature, for all spectra during the 130 s episode.  The $kT_{\rm low}$ and $kT_{\rm high}$ of the peaks represented by BB+BB models are negatively correlated with the emitting area ($r_s = -0.79$, $P < 10^{-20}$). In the same figure, we also show as a reference the luminosity contours calculated with the Stefan-Boltzmann law, assuming a distance to the source of 9 kpc. It is clear that the luminosity values remain below $\sim10^{41}$\,erg\,s$^{-1}$, which corresponds to the upper limit of magnetar short-burst luminosity.
We also note that the distribution of $kT_{\rm low}$ in the burst forest episode is narrower than the that of other short bursts from the same active episode presented in \citet{lin20}.

\subsection{\segtwo Peaks}

We expand below on the intriguing properties of Segment 2. Although the overall lightcurve of the burst forest by itself does not clearly exhibit periodic behavior consistent with the source's spin, $P_{\rm spin} =3.24$\,s, the source spectral properties are correlated with the spin period and the extrapolated pulse profile of the persistent emission observed with NICER at energies below 10 keV.  The high-energy spectral component is clearly present in both COMPT and BB+BB fits  within the burst peaks. We note that all transition times from BB+BB to the single BB thermal spectra correspond to a certain spin phase. In addition, the spectral hardness of the peaks is highest at the minima of the pulse phase, while the spectral parameters remain approximately constant during the trough intervals.  

This strong correlation between the flux and the burst spectra, as well as the phases of soft X-ray emission, provides clues to the emission and viewing geometry for this magnetar.  Because of the need to sustain bursts of duration 0.1\,s for many light-crossing times ($< 0.3$ms), it is widely believed that the bursts emanate from closed field line regions in the magnetosphere.  The extrapolated NICER lightcurve represents the persistent emission from the stellar surface that likely emanates from locations relatively near the magnetic poles \citep{younn20}. This is then more visible to an observer whose instantaneous perspective is more or less over the pole.  As the high X-ray luminosity of the forest of bursts likely indicates highly optically thick conditions (e.g., \citealt{lin11}), their zones act to occult the surface emission at various rotational phases. Thus, it is highly likely that the soft X-ray pulse peak corresponds to a phase where the observer line of sight lies roughly over the poles; this is when the GBM burst spectra are at their softest in both COMPT and BB+BB fits.  Yet, because the $<10\;$keV pulse fraction is modest \citep{younn20}, the quasi-polar surface locales are probably at least partially visible for all burst phases.  The key stellar parameters are the angle $\alpha$ between the magnetic and rotation axes, and the angle $\zeta$ between the observer direction and the spin axis.  The soft X-ray pulse profile is most easily interpreted if $\alpha$ and $\zeta$ are both not too large, likely both less than around $50^{\circ}$; more precise determination of these angles requires detailed surface modeling \citep{Younes-2020-ApJ-J1708} and treatment of magnetospheric occultation, which is beyond the scope of this Letter.

The phase dependence of the burst spectra can then be interpreted, and constraints on the magnetar geometry can be discerned.  Presume first that the various zones of burst activation span a full toroidal closed field line volume, with little or no bias in magnetic azimuth for the overall burst ensemble.  Further, suppose that the hotter regions constitute a toroidal ``belt'' at low altitudes close to where the footpoints of the field lines (that define the toroidal surface) touch the stellar surface. According to Figure\,\ref{fig:r2_kt}, this belt is roughly of effective surface area 3--30 km$^2$, a factor of 10--80 smaller than the cooler majority of the burst activation zone. For the rotational phases during which the polar surface is approximately visible to an observer, the hot zones are also readily visible along with large portions of the cooler outer toroid.  There would also be other rotational phases where the observer's view to the hotter burst locales is occulted by the optically thick cooler outer toroidal zones; these putatively correspond to the low flux phases evident in Figure\,\ref{fig:bbbb_kt_2} where the burst emission is cooler.  Given that the hotter BB component is visible for around 60--70\% of phases, the observer likely does not ever view the magnetar very close to the magnetic equator, when the hotter belt would be occulted by the cooler outer zones for larger portions of the rotation period. This translates to $\zeta$ not being very close to $90^{\circ}$, again probably less than $60^{\circ}$ or so.

A contrasting scenario would be to presume that the hotter BB regions are not adjacent to the surface but rather at higher altitudes, perhaps near the outermost portion of the toroid.  In this circumstance, because Figures~\ref{fig:flux_rad_segs} and \ref{fig:r2_kt} indicate that the areas of the hotter zones for various bursts are on average a factor 20 or so smaller than those for the cooler BB spectral components, the hotter blackbody regions would generally be obscured by the larger opaque toroid for most rotational phases.  This is contrary to the phase-dependent spectral landscape depicted in Figure\,\ref{fig:bbbb_kt_2}. Yet, it is still possible to construct pathological morphology using disparate hotspot locales -- for instance, placing the hotter BB zones on opposite faces (``upper'' and ``lower'') of the toroid and well removed from the magnetic equator.  These would have to be activated in a geometrically non-random manner at preferred magnetic longitudes in order to maintain the presence of hotter BB components for the majority of rotational phases, which would considerably constrain physical mechanisms for burst activation.

This sketch of a geometrical interpretation suggests that locales for the hotter burst zones that are well removed from the surface are somewhat disfavored. Such locales are the preserve of the magnetic reconnection scenario for magnetar burst activation \citep{Lyutikov-2003-MNRAS} that serves as an analog of the picture for solar coronal mass ejections.  They could still be accommodated, however, if the reconnection regions putatively triggering bursts do not sample magnetic longitudes in a uniform manner.  In contrast, a hot near-surface belt for the active toroids is easier to render consistent with the data in Figures~\ref{fig:bbbb_kt_2}, \ref{fig:flux_rad_segs} and \ref{fig:r2_kt}.  This is the expected locale for ignition of bursts by magnetar crustal fractures \citep{TD95,TD96}, likely involving hydromagnetic reconfigurations, and adjacent magnetospheric radiative dissipation. To discern between these two pictures requires more precise modeling of the emission geometry and its phase-dependent spectral signatures. The high quality of this burst storm dataset provides the basis for such a study, eliciting the prospect of discrimination between these burst activation scenarios.

Finally, compared with the COMPT-fit parameters presented in \citet{lin20}, the spectra of the broad peaks (at the end of \segone and the triple peaks in Segment 2) have harder $E_{\rm peak}$ and index values (Figure\,\ref{fig:compt_param}) than the time-integrated spectra of the bursts within the same active episode, observed before and after the burst forest.  However, when the time-resolved spectra of the peaks are represented with thermal components (i.e., single BB or BB+BB), the luminosity during these peaks does not exceed the upper end of the typical magnetar burst luminosity range ($\lesssim 2.6 \times 10^{41}$\,erg\,s$^{-1}$; see Figure\,\ref{fig:r2_kt}).  A further study to compare the burst forest and the time-resolved spectral properties of bursts within the same activation period is planned.

\acknowledgments
We thank Michael S. Briggs for helpful discussions regarding instrumental deadtime and pile-up issues.
Y.K and \"O.K acknowledge the support from the Scientific and Technological Research Council of Turkey (T\"{U}B\.{I}TAK grant No. 118F344).
M.G.B. acknowledges the generous support of the National Science Foundation through grant AST-1813649.

\clearpage

\begin{longrotatetable}
\begin{deluxetable*}{llcllllll}
\tablecolumns{9}
\tablenum{1}
\tablecaption{Time-resolved spectral analysis results of the Burst-forest Episode.  The spectral time resolution is 4\,ms (binned to S/N $\geq$ 25) for trough spectra and 128\,ms for peak spectra. The uncertainties are for the 1$\sigma$ confidence level.  The spectra without kT$_{\rm high}$ values are better described with a single BB either based on the BIC comparison or because the BB+BB fit failed to constrain the kT$_{\rm high}$ value.  This table is available in its entirety in the machine-readable version. \label{tab:param}}
\tablehead{
\colhead{} & \colhead{} & & \multicolumn{3}{c}{COMPT} & \multicolumn{3}{c}{BB or BB+BB} \\
\cline{4-6}
\cline{7-9}
\colhead{Start Time} & \colhead{End time} & \colhead{Flux $^{a}$} &  \colhead{E$_{\rm peak}$} & \colhead{$\Gamma$} & \colhead{C-Stat/DoF~$^{b}$}  & \colhead{kT$_{\rm low}$} & \colhead{kT$_{\rm high}$} & \colhead{C-Stat/DoF~$^{c}$}\\
\colhead{(s $- T_0^{~\dagger}$)} & \colhead{(s $- T_0^{~\dagger}$)} & \colhead{$( 10^{-6}~\rm{erg~cm}^{-2}~s^{-1})$} & \colhead{(keV)} & \colhead{} & \colhead{} & \colhead{(keV)} & \colhead{(keV)} & \colhead{}
}
\input{table1v2.tex}
\tablecomments{$^\dagger$ $T_0$ = 2020 April 27, 18:26:20 UT (609704785.155 Fermi MET) \\
$^a$ Fluence in $8-200$~keV. \\
$^b$ C-Stat for the COMPT model fit per degree of freedom (DoF). \\
$^c$ C-Stat for the BB+BB model fit or BB fit.\\
$^d$ Non-peak time bin with a low total S/N; used only for constraining the BB $kT$ and the flux.}
\end{deluxetable*}
\end{longrotatetable}

\bibliography{refs} 

\end{document}

%% file: table1v2.tex
\startdata
\multicolumn{9}{c}{Segment 1} \\
\cline{1-9}
309.89 & 316.80 & $  0.18 \pm   0.01$ & $ 27.86 \pm   0.96$ & $ -0.03 \pm   0.24$ & 172.61/259 & $   5.61 \pm   0.31$ & $ 15.21 \pm   2.00$ & 172.61/258 \\
316.80 & 316.93 & $  0.80 \pm   0.05$ & $ 24.21 \pm   2.31$ & $ -0.34 \pm   0.52$ & 283.88/259 & $   4.75 \pm   0.54$ & $ 12.97 \pm   3.10$ & 278.39/258 \\
316.93 & 317.06 & $ 19.14 \pm   0.21$ & $ 35.10 \pm   0.28$ & $  0.83 \pm   0.08$ & 297.48/259 & $   7.18 \pm   0.39$ & $ 12.12 \pm   0.72$ & 281.93/258 \\
317.06 & 317.18 & $  2.76 \pm   0.08$ & $ 32.83 \pm   0.90$ & $  0.26 \pm   0.23$ & 244.22/259 & $   5.76 \pm   0.65$ & $ 11.78 \pm   1.18$ & 241.13/258 \\
317.18 & 317.31 & $  1.64 \pm   0.07$ & $ 23.40 \pm   1.40$ & $ -0.32 \pm   0.32$ & 278.94/259 & $   5.10 \pm   0.36$ & $ 11.90 \pm   1.07$ & 277.24/258 \\
317.30 & 317.75 & $  0.66 \pm   0.03$ & $ 22.80 \pm   2.74$ & $ -1.17 \pm   0.25$ & 277.29/259 & $   4.63 \pm   0.27$ & $ 15.31 \pm   1.80$ & 277.29/258 \\
317.75 & 318.17 & $  0.78 \pm   0.03$ & $ 22.26 \pm   1.11$ & $ -0.18 \pm   0.30$ & 362.08/259 & $   5.92 \pm   0.17$ &  \nodata & 325.54/260 \\
318.17 & 319.67 & $  0.39 \pm   0.01$ & $ 22.11 \pm   1.37$ & $ -0.66 \pm   0.25$ & 289.79/259 & $   4.58 \pm   0.25$ & $ 12.92 \pm   1.61$ & 289.79/258 \\
319.67 & 321.09 & $  0.47 \pm   0.02$ & $ 42.08 \pm   1.92$ & $ -0.79 \pm   0.17$ & 350.03/259 & $   4.72 \pm   0.32$ & $ 16.00 \pm   0.89$ & 350.03/258 \\
321.02 & 321.15 & $ 23.09 \pm   0.25$ & $ 51.91 \pm   0.48$ & $  0.18 \pm   0.06$ & 355.13/259 & $   8.22 \pm   0.33$ & $ 17.80 \pm   0.49$ & 359.59/258 \\
321.15 & 321.28 & $ 29.28 \pm   0.28$ & $ 48.03 \pm   0.37$ & $  0.26 \pm   0.05$ & 326.68/259 & $   8.87 \pm   0.25$ & $ 18.55 \pm   0.64$ & 283.84/258 \\
321.28 & 328.32~$^d$ & $  0.10 \pm   0.04$ & \nodata &    \nodata &   \nodata & $  7.57 \pm   0.26$ &  \nodata & 387.92/260 \\
328.32 & 328.45 & $  0.74 \pm   0.06$ & $ 28.30 \pm   2.67$ & $ -0.44 \pm   0.50$ & 285.00/259 & $   6.32 \pm   0.22$ & $ 13.41 \pm   0.84$ & 278.86/258 \\
328.45 & 328.58 & $  4.67 \pm   0.11$ & $ 31.47 \pm   0.75$ & $ -0.09 \pm   0.16$ & 289.85/259 & $   5.32 \pm   0.36$ & $ 11.95 \pm   0.72$ & 284.64/258 \\
328.58 & 328.70 & $  9.42 \pm   0.14$ & $ 31.39 \pm   0.43$ & $  0.39 \pm   0.12$ & 290.55/259 & $   6.10 \pm   0.31$ & $ 11.93 \pm   0.72$ & 272.64/258 \\
328.70 & 328.83 & $  5.21 \pm   0.11$ & $ 29.82 \pm   0.69$ & $ -0.06 \pm   0.15$ & 335.63/259 & $   5.36 \pm   0.31$ & $ 11.89 \pm   0.72$ & 327.32/258 \\
328.83 & 328.96 & $  0.43 \pm   0.05$ & $ 22.34 \pm   4.65$ & $ -0.80 \pm   0.73$ & 289.44/259 & $   6.52 \pm   0.65$ &  \nodata & 302.84/260 \\
381.44 & 381.57 & $  8.72 \pm   0.15$ & $ 36.28 \pm   0.47$ & $  0.82 \pm   0.14$ & 273.63/259 & $   8.02 \pm   0.47$ & $ 14.55 \pm   1.90$ & 264.75/258 \\
381.57 & 381.70 & $ 18.48 \pm   0.22$ & $ 37.60 \pm   0.34$ & $  0.63 \pm   0.09$ & 293.19/259 & $   7.34 \pm   0.42$ & $ 13.08 \pm   0.70$ & 282.70/258 \\
381.70 & 381.82 & $ 20.43 \pm   0.23$ & $ 38.28 \pm   0.32$ & $  0.72 \pm   0.08$ & 305.04/259 & $   7.62 \pm   0.44$ & $ 13.22 \pm   0.76$ & 292.93/258 \\
381.82 & 381.95 & $ 22.33 \pm   0.25$ & $ 38.69 \pm   0.31$ & $  0.80 \pm   0.08$ & 315.71/259 & $   9.03 \pm   0.23$ & $ 17.27 \pm   1.75$ & 307.74/258 \\
381.95 & 382.08 & $ 27.95 \pm   0.28$ & $ 39.72 \pm   0.29$ & $  0.75 \pm   0.07$ & 294.56/259 & $   7.83 \pm   0.43$ & $ 13.31 \pm   0.62$ & 288.10/258 \\
382.08 & 382.21 & $ 30.63 \pm   0.29$ & $ 39.38 \pm   0.27$ & $  0.79 \pm   0.07$ & 295.74/259 & $   8.01 \pm   0.37$ & $ 13.69 \pm   0.69$ & 276.24/258 \\
382.21 & 382.34 & $ 34.32 \pm   0.31$ & $ 40.11 \pm   0.26$ & $  0.72 \pm   0.06$ & 291.05/259 & $   7.09 \pm   0.43$ & $ 12.57 \pm   0.37$ & 289.75/258 \\
382.34 & 382.46 & $ 35.58 \pm   0.32$ & $ 40.50 \pm   0.27$ & $  0.68 \pm   0.06$ & 280.08/259 & $   7.88 \pm   0.36$ & $ 13.74 \pm   0.54$ & 270.22/258 \\
382.46 & 382.59 & $ 36.40 \pm   0.32$ & $ 41.04 \pm   0.27$ & $  0.60 \pm   0.06$ & 286.91/259 & $   7.72 \pm   0.33$ & $ 13.98 \pm   0.48$ & 268.96/258 \\
382.59 & 382.72 & $ 35.72 \pm   0.32$ & $ 40.97 \pm   0.26$ & $  0.73 \pm   0.06$ & 326.94/259 & $   8.26 \pm   0.36$ & $ 14.16 \pm   0.64$ & 313.74/258 \\
382.72 & 382.85 & $ 35.77 \pm   0.32$ & $ 40.90 \pm   0.26$ & $  0.74 \pm   0.06$ & 291.64/259 & $   8.42 \pm   0.34$ & $ 14.51 \pm   0.72$ & 275.31/258 \\
382.85 & 382.98 & $ 36.18 \pm   0.32$ & $ 41.15 \pm   0.27$ & $  0.62 \pm   0.06$ & 328.97/259 & $   8.65 \pm   0.26$ & $ 15.76 \pm   0.79$ & 302.69/258 \\
382.98 & 383.10 & $ 35.56 \pm   0.32$ & $ 41.14 \pm   0.28$ & $  0.58 \pm   0.06$ & 287.83/259 & $   8.06 \pm   0.31$ & $ 14.58 \pm   0.58$ & 269.54/258 \\
383.10 & 383.23 & $ 36.87 \pm   0.33$ & $ 42.49 \pm   0.28$ & $  0.54 \pm   0.06$ & 328.62/259 & $   8.22 \pm   0.29$ & $ 15.27 \pm   0.57$ & 297.51/258 \\
383.23 & 383.36 & $ 37.52 \pm   0.33$ & $ 41.72 \pm   0.27$ & $  0.68 \pm   0.06$ & 359.27/259 & $   8.56 \pm   0.28$ & $ 15.51 \pm   0.69$ & 315.55/258 \\
383.36 & 383.49 & $ 34.68 \pm   0.31$ & $ 40.97 \pm   0.27$ & $  0.68 \pm   0.06$ & 315.20/259 & $   8.21 \pm   0.33$ & $ 14.48 \pm   0.64$ & 295.12/258 \\
383.49 & 383.62 & $ 19.65 \pm   0.23$ & $ 35.67 \pm   0.33$ & $  0.49 \pm   0.08$ & 290.52/259 & $   6.58 \pm   0.33$ & $ 12.38 \pm   0.49$ & 276.85/258 \\
383.60 & 391.90 & $  0.02 \pm   0.01$ & $ 17.64 \pm   8.47$ & $ -1.16 \pm   1.01$ & 215.79/259 & $   5.13 \pm   0.66$ &  \nodata & 215.96/260 \\
\multicolumn{9}{c}{Segment 2} \\
\cline{1-9}
391.94 & 394.82 & $  0.14 \pm   0.01$ & $ 27.24 \pm   1.46$ & $  0.24 \pm   0.42$ & 368.76/346 & $   7.07 \pm   0.27$ &  \nodata & 379.24/347 \\
394.82 & 394.85 & $  8.71 \pm   0.25$ & $ 28.22 \pm   0.77$ & $  0.46 \pm   0.23$ & 374.77/346 & $   7.54 \pm   0.13$ &  \nodata & 398.05/347 \\
394.85 & 396.14 & $  0.30 \pm   0.01$ & $ 24.98 \pm   1.02$ & $  0.62 \pm   0.33$ & 475.13/346 & $   6.40 \pm   0.19$ &  \nodata & 486.64/347 \\
396.14 & 399.83 & $  0.11 \pm   0.01$ & $ 20.06 \pm   3.99$ & $  0.53 \pm   0.36$ & 409.23/346 & $   5.81 \pm   0.29$ &  \nodata & 443.07/347 \\
399.83 & 402.94 & $  0.15 \pm   0.01$ & $ 11.41 \pm  11.14$ & $  0.50 \pm   0.25$ & 355.19/346 & $   5.23 \pm   0.25$ &  \nodata & 404.06/347 \\
402.94 & 402.98 & $  7.09 \pm   0.20$ & $ 33.00 \pm   1.07$ & $  0.41 \pm   0.18$ & 338.31/346 & $   5.42 \pm   0.39$ & $ 13.17 \pm   0.82$ & 337.40/345 \\
402.98 & 403.02 & $  8.09 \pm   0.23$ & $ 31.43 \pm   0.90$ & $  0.20 \pm   0.20$ & 339.57/346 & $   6.03 \pm   0.45$ & $ 13.26 \pm   1.34$ & 326.57/345 \\
403.02 & 404.76 & $  0.22 \pm   0.01$ & $ 24.90 \pm   1.93$ & $  0.19 \pm   0.31$ & 424.21/346 & $   6.40 \pm   0.23$ &  \nodata & 501.58/347 \\
404.76 & 404.78 & $ 14.79 \pm   0.40$ & $ 33.79 \pm   0.72$ & $  0.22 \pm   0.21$ & 302.72/346 & $   8.02 \pm   0.43$ & $ 16.69 \pm   4.67$ & 299.62/345 \\
404.78 & 404.79 & $ 18.20 \pm   0.49$ & $ 33.80 \pm   0.73$ & $  0.32 \pm   0.21$ & 291.84/346 & $   6.22 \pm   0.85$ & $ 11.32 \pm   1.10$ & 288.59/345 \\
404.79 & 404.81 & $ 16.20 \pm   0.45$ & $ 32.46 \pm   0.72$ & $  0.41 \pm   0.22$ & 273.41/346 & $   8.57 \pm   0.14$ &  \nodata & 292.71/347 \\
404.81 & 404.84 & $  8.92 \pm   0.25$ & $ 29.26 \pm   0.82$ & $  0.41 \pm   0.22$ & 314.71/346 & $   7.79 \pm   0.14$ &  \nodata & 343.33/347 \\
404.84 & 404.96 & $  2.53 \pm   0.08$ & $ 27.31 \pm   1.00$ & $  0.27 \pm   0.22$ & 432.80/346 & $   7.23 \pm   0.15$ &  \nodata & 471.23/347 \\
404.96 & 405.94 & $  0.43 \pm   0.02$ & $ 24.13 \pm   2.09$ & $  0.24 \pm   0.22$ & 377.13/346 & $   7.03 \pm   0.19$ &  \nodata & 456.41/347 \\
406.02 & 406.14 & $ 12.44 \pm   0.15$ & $ 36.32 \pm   0.38$ & $  0.39 \pm   0.08$ & 470.42/346 & $   6.28 \pm   0.28$ & $ 12.92 \pm   0.44$ & 453.96/345 \\
406.17 & 406.21 & $  6.65 \pm   0.19$ & $ 30.87 \pm   0.83$ & $  0.49 \pm   0.21$ & 338.34/346 & $   6.54 \pm   0.39$ & $ 14.88 \pm   2.04$ & 323.02/345 \\
406.21 & 406.33 & $  2.64 \pm   0.08$ & $ 24.95 \pm   1.19$ & $  0.42 \pm   0.21$ & 442.20/346 & $   7.00 \pm   0.14$ &  \nodata & 510.47/347 \\
406.33 & 406.73 & $  0.86 \pm   0.03$ & $ 19.33 \pm   2.02$ & $  0.54 \pm   0.22$ & 403.68/346 & $   5.76 \pm   0.15$ &  \nodata & 470.51/347 \\
406.73 & 406.93 & $  1.59 \pm   0.05$ & $ 27.29 \pm   1.00$ & $  0.37 \pm   0.23$ & 379.82/346 & $   7.31 \pm   0.15$ &  \nodata & 425.68/347 \\
406.93 & 406.96 & $  9.98 \pm   0.27$ & $ 31.16 \pm   0.81$ & $  0.15 \pm   0.21$ & 355.06/346 & $   5.68 \pm   0.51$ & $ 11.75 \pm   1.01$ & 345.62/345 \\
406.96 & 407.04 & $  3.52 \pm   0.10$ & $ 28.07 \pm   0.81$ & $  0.14 \pm   0.23$ & 356.98/346 & $   7.50 \pm   0.14$ &  \nodata & 388.36/347 \\
407.04 & 407.16 & $  2.53 \pm   0.07$ & $ 26.23 \pm   0.86$ & $  0.18 \pm   0.23$ & 357.62/346 & $   7.09 \pm   0.14$ &  \nodata & 391.79/347 \\
407.16 & 407.30 & $  2.25 \pm   0.07$ & $ 26.32 \pm   1.36$ & $  0.11 \pm   0.20$ & 376.07/346 & $   7.45 \pm   0.15$ &  \nodata & 470.08/347 \\
407.30 & 407.36 & $  4.55 \pm   0.13$ & $ 31.04 \pm   0.74$ & $  0.17 \pm   0.23$ & 316.53/346 & $   8.12 \pm   0.14$ &  \nodata & 336.63/347 \\
407.36 & 407.56 & $  1.54 \pm   0.05$ & $ 26.42 \pm   1.03$ & $  0.16 \pm   0.23$ & 406.16/346 & $   7.29 \pm   0.15$ &  \nodata & 453.85/347 \\
407.56 & 407.86 & $  1.04 \pm   0.03$ & $ 23.37 \pm   1.13$ & $  0.12 \pm   0.26$ & 385.86/346 & $   6.35 \pm   0.15$ &  \nodata & 426.53/347 \\
407.81 & 407.94 & $ 12.38 \pm   0.15$ & $ 36.08 \pm   0.35$ & $  0.33 \pm   0.09$ & 407.55/346 & $   6.38 \pm   0.36$ & $ 12.17 \pm   0.45$ & 399.87/345 \\
407.94 & 408.06 & $ 27.73 \pm   0.23$ & $ 39.72 \pm   0.25$ & $  0.18 \pm   0.06$ & 482.07/346 & $   7.57 \pm   0.32$ & $ 13.45 \pm   0.45$ & 471.01/345 \\
408.06 & 408.19 & $ 32.02 \pm   0.25$ & $ 41.12 \pm   0.25$ & $  0.30 \pm   0.05$ & 497.27/346 & $   7.34 \pm   0.30$ & $ 13.55 \pm   0.34$ & 501.60/345 \\
408.19 & 408.32 & $ 31.98 \pm   0.25$ & $ 42.26 \pm   0.26$ & $  0.52 \pm   0.05$ & 533.14/346 & $   7.80 \pm   0.29$ & $ 14.37 \pm   0.41$ & 536.75/345 \\
408.32 & 408.45 & $ 31.13 \pm   0.25$ & $ 43.65 \pm   0.27$ & $  0.45 \pm   0.05$ & 522.52/346 & $   7.73 \pm   0.26$ & $ 15.08 \pm   0.38$ & 502.14/345 \\
408.45 & 408.58 & $ 31.75 \pm   0.26$ & $ 44.71 \pm   0.30$ & $  0.47 \pm   0.05$ & 514.00/346 & $   7.45 \pm   0.23$ & $ 15.57 \pm   0.33$ & 518.98/345 \\
408.58 & 408.70 & $ 32.15 \pm   0.26$ & $ 45.50 \pm   0.30$ & $  0.33 \pm   0.05$ & 541.85/346 & $   8.11 \pm   0.21$ & $ 16.90 \pm   0.42$ & 512.78/345 \\
408.70 & 408.83 & $ 31.84 \pm   0.25$ & $ 44.31 \pm   0.29$ & $  0.33 \pm   0.05$ & 556.41/346 & $   7.70 \pm   0.22$ & $ 16.02 \pm   0.38$ & 516.61/345 \\
408.83 & 408.96 & $ 31.36 \pm   0.25$ & $ 42.67 \pm   0.27$ & $  0.26 \pm   0.05$ & 575.31/346 & $   7.97 \pm   0.22$ & $ 15.83 \pm   0.45$ & 543.90/345 \\
408.96 & 409.09 & $ 30.59 \pm   0.25$ & $ 40.85 \pm   0.26$ & $  0.43 \pm   0.05$ & 517.89/346 & $   7.71 \pm   0.24$ & $ 14.78 \pm   0.44$ & 490.07/345 \\
409.09 & 409.22 & $ 28.69 \pm   0.23$ & $ 38.77 \pm   0.25$ & $  0.23 \pm   0.05$ & 573.98/346 & $   7.18 \pm   0.25$ & $ 13.60 \pm   0.37$ & 568.57/345 \\
409.22 & 409.34 & $ 30.63 \pm   0.24$ & $ 38.29 \pm   0.25$ & $  0.34 \pm   0.05$ & 573.97/346 & $   7.41 \pm   0.18$ & $ 15.03 \pm   0.44$ & 504.67/345 \\
409.34 & 409.47 & $ 31.14 \pm   0.24$ & $ 35.80 \pm   0.24$ & $  0.08 \pm   0.05$ & 644.85/346 & $   6.37 \pm   0.18$ & $ 12.89 \pm   0.29$ & 614.96/345 \\
409.47 & 409.60 & $ 28.34 \pm   0.23$ & $ 35.08 \pm   0.24$ & $  0.13 \pm   0.06$ & 674.61/346 & $   6.72 \pm   0.20$ & $ 12.88 \pm   0.37$ & 638.43/345 \\
409.60 & 409.73 & $ 26.48 \pm   0.22$ & $ 33.96 \pm   0.24$ & $  0.16 \pm   0.06$ & 592.78/346 & $   6.56 \pm   0.23$ & $ 12.02 \pm   0.37$ & 578.08/345 \\
409.73 & 409.86 & $ 21.38 \pm   0.20$ & $ 35.53 \pm   0.27$ & $  0.33 \pm   0.07$ & 620.41/346 & $   6.19 \pm   0.27$ & $ 11.94 \pm   0.32$ & 612.76/345 \\
409.86 & 409.98 & $ 16.26 \pm   0.17$ & $ 35.27 \pm   0.30$ & $  0.29 \pm   0.08$ & 488.29/346 & $   7.29 \pm   0.27$ & $ 13.60 \pm   0.71$ & 461.59/345 \\
409.98 & 410.11 & $ 12.54 \pm   0.15$ & $ 34.86 \pm   0.36$ & $  0.36 \pm   0.09$ & 452.38/346 & $   7.57 \pm   0.24$ & $ 15.02 \pm   1.12$ & 456.48/345 \\
410.17 & 410.20 & $  7.48 \pm   0.21$ & $ 30.80 \pm   0.83$ & $  0.28 \pm   0.22$ & 298.47/346 & $   8.20 \pm   0.15$ &  \nodata & 325.55/347 \\
410.20 & 410.24 & $  8.27 \pm   0.23$ & $ 30.40 \pm   0.84$ & $  0.52 \pm   0.21$ & 339.03/346 & $   8.15 \pm   0.14$ &  \nodata & 375.88/347 \\
410.24 & 410.28 & $  7.36 \pm   0.20$ & $ 30.18 \pm   0.75$ & $  0.40 \pm   0.23$ & 324.99/346 & $   7.96 \pm   0.14$ &  \nodata & 346.88/347 \\
410.28 & 410.32 & $  6.97 \pm   0.19$ & $ 30.13 \pm   0.70$ & $  0.36 \pm   0.23$ & 335.08/346 & $   7.94 \pm   0.14$ &  \nodata & 347.18/347 \\
410.32 & 410.37 & $  7.08 \pm   0.19$ & $ 30.80 \pm   0.75$ & $  0.14 \pm   0.22$ & 311.73/346 & $   8.14 \pm   0.14$ &  \nodata & 334.62/347 \\
410.37 & 410.41 & $  7.41 \pm   0.20$ & $ 29.04 \pm   0.77$ & $  0.37 \pm   0.22$ & 299.43/346 & $   7.77 \pm   0.14$ &  \nodata & 326.11/347 \\
410.41 & 410.45 & $  7.08 \pm   0.19$ & $ 30.05 \pm   0.73$ & $  0.37 \pm   0.22$ & 313.36/346 & $   7.93 \pm   0.14$ &  \nodata & 335.76/347 \\
410.45 & 410.49 & $  7.08 \pm   0.20$ & $ 28.83 \pm   0.85$ & $  0.37 \pm   0.22$ & 325.26/346 & $   7.75 \pm   0.14$ &  \nodata & 360.77/347 \\
410.49 & 410.53 & $  6.67 \pm   0.19$ & $ 30.04 \pm   0.75$ & $  0.37 \pm   0.22$ & 387.14/346 & $   7.90 \pm   0.14$ &  \nodata & 406.78/347 \\
410.53 & 410.58 & $  6.05 \pm   0.17$ & $ 27.13 \pm   0.87$ & $  0.37 \pm   0.22$ & 323.02/346 & $   7.36 \pm   0.14$ &  \nodata & 356.82/347 \\
410.58 & 410.63 & $  6.34 \pm   0.17$ & $ 29.19 \pm   0.70$ & $  0.37 \pm   0.23$ & 351.34/346 & $   7.71 \pm   0.14$ &  \nodata & 365.46/347 \\
410.63 & 410.68 & $  5.39 \pm   0.15$ & $ 27.59 \pm   0.67$ & $  0.37 \pm   0.25$ & 310.34/346 & $   7.28 \pm   0.14$ &  \nodata & 321.31/347 \\
410.68 & 410.75 & $  4.26 \pm   0.12$ & $ 26.72 \pm   0.85$ & $  0.37 \pm   0.23$ & 372.07/346 & $   7.30 \pm   0.13$ &  \nodata & 411.18/347 \\
410.75 & 410.83 & $  3.95 \pm   0.11$ & $ 26.94 \pm   0.89$ & $  0.37 \pm   0.22$ & 348.91/346 & $   7.36 \pm   0.14$ &  \nodata & 390.20/347 \\
410.83 & 410.90 & $  3.58 \pm   0.11$ & $ 25.55 \pm   0.81$ & $  0.37 \pm   0.25$ & 378.08/346 & $   6.88 \pm   0.14$ &  \nodata & 400.91/347 \\
410.90 & 411.12 & $  1.47 \pm   0.05$ & $ 26.22 \pm   1.10$ & $  0.37 \pm   0.23$ & 436.05/346 & $   7.24 \pm   0.15$ &  \nodata & 484.25/347 \\
411.12 & 411.28 & $  2.00 \pm   0.06$ & $ 26.73 \pm   1.02$ & $  0.37 \pm   0.23$ & 442.98/346 & $   7.21 \pm   0.14$ &  \nodata & 490.33/347 \\
411.28 & 411.35 & $  4.63 \pm   0.13$ & $ 31.44 \pm   0.94$ & $  0.37 \pm   0.20$ & 350.00/346 & $   8.36 \pm   0.15$ &  \nodata & 393.60/347 \\
411.35 & 411.38 & $ 11.10 \pm   0.30$ & $ 34.89 \pm   0.69$ & $  0.37 \pm   0.22$ & 324.28/346 & $   9.08 \pm   0.15$ &  \nodata & 331.20/347 \\
411.38 & 411.42 & $  6.16 \pm   0.18$ & $ 31.30 \pm   0.98$ & $  0.37 \pm   0.20$ & 375.09/346 & $   5.76 \pm   0.42$ & $ 13.19 \pm   1.13$ & 365.15/345 \\
411.42 & 411.53 & $  3.02 \pm   0.09$ & $ 27.72 \pm   1.06$ & $  0.37 \pm   0.21$ & 399.40/346 & $   7.52 \pm   0.14$ &  \nodata & 460.07/347 \\
411.53 & 411.58 & $  5.63 \pm   0.16$ & $ 32.66 \pm   0.96$ & $  0.37 \pm   0.19$ & 327.08/346 & $   5.30 \pm   0.45$ & $ 12.14 \pm   0.80$ & 322.68/345 \\
411.58 & 411.60 & $ 22.65 \pm   0.61$ & $ 39.45 \pm   0.90$ & $  0.37 \pm   0.17$ & 278.73/346 & $   6.23 \pm   0.65$ & $ 13.53 \pm   0.82$ & 276.76/345 \\
411.60 & 411.62 & $ 20.57 \pm   0.55$ & $ 38.34 \pm   0.95$ & $  0.37 \pm   0.17$ & 303.88/346 & $   6.07 \pm   0.52$ & $ 14.03 \pm   0.82$ & 297.43/345 \\
411.65 & 411.78 & $ 13.87 \pm   0.16$ & $ 34.63 \pm   0.37$ & $  0.37 \pm   0.08$ & 469.39/346 & $   6.57 \pm   0.23$ & $ 13.76 \pm   0.53$ & 435.41/345 \\
411.78 & 411.90 & $ 21.26 \pm   0.20$ & $ 34.85 \pm   0.30$ & $  0.37 \pm   0.06$ & 516.24/346 & $   6.88 \pm   0.18$ & $ 14.48 \pm   0.49$ & 462.89/345 \\
411.90 & 412.03 & $ 23.64 \pm   0.21$ & $ 37.22 \pm   0.29$ & $  0.37 \pm   0.06$ & 502.32/346 & $   6.90 \pm   0.20$ & $ 14.11 \pm   0.40$ & 480.75/345 \\
412.03 & 412.16 & $ 21.28 \pm   0.20$ & $ 37.14 \pm   0.31$ & $  0.37 \pm   0.06$ & 457.10/346 & $   6.73 \pm   0.20$ & $ 14.21 \pm   0.40$ & 424.06/345 \\
412.16 & 412.29 & $ 28.10 \pm   0.24$ & $ 39.68 \pm   0.28$ & $  0.37 \pm   0.05$ & 572.91/346 & $   7.33 \pm   0.19$ & $ 15.24 \pm   0.41$ & 530.23/345 \\
412.29 & 412.42 & $ 36.77 \pm   0.27$ & $ 40.98 \pm   0.26$ & $  0.37 \pm   0.05$ & 605.14/346 & $   6.96 \pm   0.19$ & $ 14.56 \pm   0.28$ & 598.95/345 \\
412.42 & 412.54 & $ 40.45 \pm   0.29$ & $ 41.21 \pm   0.25$ & $  0.37 \pm   0.04$ & 557.43/346 & $   7.17 \pm   0.17$ & $ 15.09 \pm   0.29$ & 550.29/345 \\
412.54 & 412.67 & $ 36.20 \pm   0.27$ & $ 39.08 \pm   0.24$ & $  0.37 \pm   0.05$ & 612.15/346 & $   6.78 \pm   0.21$ & $ 13.55 \pm   0.28$ & 584.27/345 \\
412.67 & 412.80 & $ 40.03 \pm   0.29$ & $ 40.30 \pm   0.24$ & $  0.37 \pm   0.04$ & 557.42/346 & $   7.02 \pm   0.19$ & $ 14.34 \pm   0.28$ & 573.98/345 \\
412.80 & 412.93 & $ 38.07 \pm   0.28$ & $ 39.35 \pm   0.24$ & $  0.37 \pm   0.05$ & 612.82/346 & $   6.67 \pm   0.21$ & $ 13.40 \pm   0.25$ & 617.48/345 \\
412.93 & 413.06 & $ 35.07 \pm   0.26$ & $ 38.79 \pm   0.23$ & $  0.37 \pm   0.05$ & 585.98/346 & $   8.00 \pm   0.20$ & $ 15.04 \pm   0.55$ & 529.34/345 \\
413.06 & 413.18 & $ 25.22 \pm   0.22$ & $ 38.00 \pm   0.26$ & $  0.37 \pm   0.06$ & 462.28/346 & $   6.85 \pm   0.27$ & $ 13.02 \pm   0.36$ & 445.41/345 \\
413.18 & 413.31 & $ 14.27 \pm   0.16$ & $ 35.08 \pm   0.33$ & $  0.37 \pm   0.08$ & 494.72/346 & $   7.26 \pm   0.25$ & $ 14.14 \pm   0.77$ & 460.35/345 \\
413.31 & 413.44 & $  9.84 \pm   0.13$ & $ 33.61 \pm   0.40$ & $  0.37 \pm   0.10$ & 409.12/346 & $   6.17 \pm   0.33$ & $ 11.94 \pm   0.52$ & 405.81/345 \\
413.44 & 413.57 & $  7.48 \pm   0.12$ & $ 32.06 \pm   0.46$ & $  0.37 \pm   0.12$ & 395.97/346 & $   8.16 \pm   0.09$ & $ 21.18 \pm   2.69$ & 421.54/346 \\
413.57 & 413.61 & $  7.35 \pm   0.21$ & $ 30.58 \pm   0.82$ & $  0.37 \pm   0.22$ & 306.38/346 & $   8.21 \pm   0.14$ &  \nodata & 344.22/347 \\
413.61 & 413.65 & $  7.19 \pm   0.20$ & $ 30.12 \pm   0.81$ & $  0.37 \pm   0.21$ & 329.21/346 & $   8.05 \pm   0.14$ &  \nodata & 359.36/347 \\
413.65 & 413.69 & $  7.66 \pm   0.21$ & $ 29.66 \pm   0.77$ & $  0.37 \pm   0.22$ & 316.21/346 & $   7.81 \pm   0.14$ &  \nodata & 338.56/347 \\
413.69 & 413.73 & $  7.89 \pm   0.22$ & $ 30.34 \pm   0.73$ & $  0.37 \pm   0.22$ & 312.40/346 & $   8.00 \pm   0.14$ &  \nodata & 331.55/347 \\
413.73 & 413.77 & $  8.18 \pm   0.22$ & $ 31.24 \pm   0.87$ & $  0.37 \pm   0.20$ & 372.57/346 & $   6.68 \pm   0.33$ & $ 16.83 \pm   2.10$ & 344.53/345 \\
413.77 & 413.80 & $  7.60 \pm   0.21$ & $ 30.53 \pm   0.87$ & $  0.37 \pm   0.21$ & 333.53/346 & $   8.24 \pm   0.14$ &  \nodata & 379.40/347 \\
413.80 & 413.84 & $  7.84 \pm   0.22$ & $ 30.99 \pm   0.84$ & $  0.37 \pm   0.20$ & 350.78/346 & $   7.28 \pm   0.24$ & $ 22.91 \pm   3.61$ & 328.68/345 \\
413.84 & 413.89 & $  7.24 \pm   0.20$ & $ 30.84 \pm   0.79$ & $  0.37 \pm   0.21$ & 295.97/346 & $   8.18 \pm   0.14$ &  \nodata & 323.89/347 \\
413.89 & 413.93 & $  7.42 \pm   0.20$ & $ 30.42 \pm   0.69$ & $  0.37 \pm   0.23$ & 303.20/346 & $   8.01 \pm   0.14$ &  \nodata & 316.06/347 \\
413.93 & 413.97 & $  7.59 \pm   0.21$ & $ 32.79 \pm   0.86$ & $  0.37 \pm   0.20$ & 343.29/346 & $   7.02 \pm   0.41$ & $ 15.96 \pm   2.16$ & 331.24/345 \\
413.97 & 414.01 & $  7.68 \pm   0.21$ & $ 31.54 \pm   0.76$ & $  0.37 \pm   0.21$ & 321.76/346 & $   8.26 \pm   0.14$ &  \nodata & 344.88/347 \\
414.01 & 414.05 & $  7.77 \pm   0.21$ & $ 32.55 \pm   0.65$ & $  0.37 \pm   0.24$ & 319.91/346 & $   8.46 \pm   0.15$ &  \nodata & 325.13/347 \\
414.05 & 414.09 & $  7.36 \pm   0.20$ & $ 30.35 \pm   0.79$ & $  0.37 \pm   0.20$ & 332.70/346 & $   8.09 \pm   0.14$ &  \nodata & 367.93/347 \\
414.09 & 414.14 & $  7.05 \pm   0.19$ & $ 31.82 \pm   0.78$ & $  0.37 \pm   0.21$ & 314.30/346 & $   8.38 \pm   0.14$ &  \nodata & 339.80/347 \\
414.14 & 414.18 & $  7.59 \pm   0.21$ & $ 30.18 \pm   0.85$ & $  0.37 \pm   0.20$ & 300.37/346 & $   8.11 \pm   0.14$ &  \nodata & 341.75/347 \\
414.18 & 414.21 & $  8.01 \pm   0.22$ & $ 32.01 \pm   0.73$ & $  0.37 \pm   0.22$ & 313.18/346 & $   8.37 \pm   0.14$ &  \nodata & 331.14/347 \\
414.21 & 414.25 & $  8.56 \pm   0.24$ & $ 33.00 \pm   0.82$ & $  0.37 \pm   0.21$ & 356.29/346 & $   6.91 \pm   0.55$ & $ 14.04 \pm   1.98$ & 348.95/345 \\
414.25 & 414.29 & $  7.75 \pm   0.21$ & $ 31.95 \pm   0.79$ & $  0.37 \pm   0.21$ & 322.70/346 & $   6.49 \pm   0.53$ & $ 12.96 \pm   1.50$ & 317.63/345 \\
414.29 & 414.33 & $  7.13 \pm   0.20$ & $ 31.37 \pm   0.71$ & $  0.37 \pm   0.23$ & 345.52/346 & $   8.23 \pm   0.14$ &  \nodata & 360.95/347 \\
414.33 & 414.37 & $  8.46 \pm   0.23$ & $ 31.96 \pm   0.78$ & $  0.37 \pm   0.21$ & 302.98/346 & $   8.46 \pm   0.14$ &  \nodata & 333.10/347 \\
414.37 & 414.41 & $  8.36 \pm   0.23$ & $ 32.34 \pm   0.77$ & $  0.37 \pm   0.21$ & 307.77/346 & $   8.53 \pm   0.14$ &  \nodata & 335.66/347 \\
414.41 & 414.44 & $  8.73 \pm   0.23$ & $ 32.55 \pm   0.67$ & $  0.37 \pm   0.22$ & 303.90/346 & $   8.51 \pm   0.14$ &  \nodata & 315.13/347 \\
414.44 & 414.48 & $  8.62 \pm   0.25$ & $ 32.70 \pm   0.83$ & $  0.37 \pm   0.21$ & 337.06/346 & $   6.76 \pm   0.52$ & $ 14.01 \pm   1.84$ & 328.14/345 \\
414.48 & 414.51 & $  8.72 \pm   0.24$ & $ 32.82 \pm   0.86$ & $  0.37 \pm   0.20$ & 305.01/346 & $   6.12 \pm   0.51$ & $ 13.03 \pm   1.22$ & 293.90/345 \\
414.59 & 414.72 & $ 10.07 \pm   0.14$ & $ 36.03 \pm   0.42$ & $  0.37 \pm   0.10$ & 385.60/346 & $   6.31 \pm   0.34$ & $ 12.75 \pm   0.52$ & 377.85/345 \\
414.72 & 414.85 & $ 14.18 \pm   0.16$ & $ 38.32 \pm   0.35$ & $  0.37 \pm   0.08$ & 448.40/346 & $   6.40 \pm   0.36$ & $ 12.64 \pm   0.39$ & 441.41/345 \\
414.85 & 414.98 & $ 11.41 \pm   0.15$ & $ 36.40 \pm   0.41$ & $  0.37 \pm   0.09$ & 437.28/346 & $   6.91 \pm   0.28$ & $ 14.14 \pm   0.64$ & 413.98/345 \\
414.98 & 415.10 & $ 10.21 \pm   0.14$ & $ 35.88 \pm   0.41$ & $  0.37 \pm   0.10$ & 438.86/346 & $   7.02 \pm   0.31$ & $ 13.94 \pm   0.75$ & 418.01/345 \\
415.10 & 415.23 & $ 10.32 \pm   0.14$ & $ 38.91 \pm   0.47$ & $  0.37 \pm   0.09$ & 406.91/346 & $   6.49 \pm   0.30$ & $ 14.16 \pm   0.51$ & 391.50/345 \\
415.23 & 415.36 & $ 10.70 \pm   0.14$ & $ 35.80 \pm   0.43$ & $  0.37 \pm   0.09$ & 398.61/346 & $   6.80 \pm   0.26$ & $ 14.37 \pm   0.64$ & 376.29/345 \\
415.36 & 415.49 & $ 21.03 \pm   0.20$ & $ 36.04 \pm   0.30$ & $  0.37 \pm   0.06$ & 568.99/346 & $   6.37 \pm   0.21$ & $ 13.21 \pm   0.34$ & 549.68/345 \\
415.49 & 415.62 & $ 32.59 \pm   0.25$ & $ 36.63 \pm   0.24$ & $  0.37 \pm   0.05$ & 546.36/346 & $   7.24 \pm   0.18$ & $ 14.29 \pm   0.42$ & 485.07/345 \\
415.62 & 415.74 & $ 30.05 \pm   0.24$ & $ 35.81 \pm   0.25$ & $  0.37 \pm   0.05$ & 568.77/346 & $   6.99 \pm   0.18$ & $ 13.81 \pm   0.39$ & 527.42/345 \\
415.74 & 415.87 & $ 28.17 \pm   0.23$ & $ 34.70 \pm   0.24$ & $  0.37 \pm   0.06$ & 564.02/346 & $   7.31 \pm   0.16$ & $ 14.70 \pm   0.55$ & 493.70/345 \\
415.87 & 416.00 & $ 24.61 \pm   0.21$ & $ 34.62 \pm   0.26$ & $  0.37 \pm   0.06$ & 469.65/346 & $   7.03 \pm   0.17$ & $ 14.19 \pm   0.51$ & 412.31/345 \\
416.00 & 416.13 & $ 21.56 \pm   0.20$ & $ 34.51 \pm   0.26$ & $  0.37 \pm   0.07$ & 548.36/346 & $   6.90 \pm   0.24$ & $ 12.74 \pm   0.50$ & 518.01/345 \\
416.13 & 416.26 & $ 19.93 \pm   0.19$ & $ 33.08 \pm   0.27$ & $  0.37 \pm   0.07$ & 492.50/346 & $  6.05 \pm   0.23$ & $ 11.65 \pm   0.35$ & 467.82/345 \\
416.26 & 416.38 & $ 13.25 \pm   0.15$ & $ 32.35 \pm   0.34$ & $  0.37 \pm   0.09$ & 415.50/346 & $   6.26 \pm   0.26$ & $ 12.16 \pm   0.52$ & 391.88/345 \\
416.39 & 416.41 & $ 12.83 \pm   0.35$ & $ 32.87 \pm   0.71$ & $  0.37 \pm   0.22$ & 261.99/346 & $   8.61 \pm   0.14$ &  \nodata & 279.43/347 \\
416.41 & 416.43 & $ 14.93 \pm   0.40$ & $ 33.57 \pm   0.73$ & $  0.37 \pm   0.20$ & 264.12/346 & $   6.22 \pm   0.79$ & $ 11.37 \pm   1.06$ & 261.96/345 \\
416.43 & 416.45 & $ 16.61 \pm   0.44$ & $ 31.55 \pm   0.72$ & $  0.37 \pm   0.21$ & 278.28/346 & $   8.40 \pm   0.13$ &  \nodata & 305.75/347 \\
416.45 & 416.47 & $ 17.08 \pm   0.46$ & $ 31.32 \pm   0.69$ & $  0.37 \pm   0.22$ & 267.14/346 & $   8.27 \pm   0.14$ &  \nodata & 285.09/347 \\
416.47 & 416.49 & $ 17.15 \pm   0.45$ & $ 30.49 \pm   0.71$ & $  0.37 \pm   0.21$ & 274.56/346 & $   8.09 \pm   0.13$ &  \nodata & 296.50/347 \\
416.49 & 416.51 & $ 16.32 \pm   0.45$ & $ 31.11 \pm   0.69$ & $  0.37 \pm   0.23$ & 263.68/346 & $   8.21 \pm   0.14$ &  \nodata & 279.52/347 \\
416.51 & 416.53 & $ 17.52 \pm   0.48$ & $ 31.43 \pm   0.71$ & $  0.37 \pm   0.22$ & 243.77/346 & $   8.30 \pm   0.14$ &  \nodata & 261.25/347 \\
416.53 & 416.54 & $ 18.37 \pm   0.50$ & $ 32.42 \pm   0.76$ & $  0.37 \pm   0.20$ & 263.81/346 & $   8.60 \pm   0.14$ &  \nodata & 289.78/347 \\
416.54 & 416.56 & $ 16.38 \pm   0.45$ & $ 31.32 \pm   0.79$ & $  0.37 \pm   0.21$ & 304.69/346 & $   6.57 \pm   0.47$ & $ 13.47 \pm   1.81$ & 293.84/345 \\
416.56 & 416.58 & $ 16.37 \pm   0.44$ & $ 31.49 \pm   0.76$ & $  0.37 \pm   0.21$ & 287.68/346 & $   7.06 \pm   0.40$ & $ 15.07 \pm   2.29$ & 278.62/345 \\
416.58 & 416.60 & $ 14.60 \pm   0.40$ & $ 29.79 \pm   0.78$ & $  0.37 \pm   0.21$ & 320.38/346 & $   7.98 \pm   0.13$ &  \nodata & 351.55/347 \\
416.60 & 416.62 & $ 14.77 \pm   0.39$ & $ 30.02 \pm   0.72$ & $  0.37 \pm   0.21$ & 282.36/346 & $   7.99 \pm   0.13$ &  \nodata & 308.11/347 \\
416.62 & 416.65 & $ 13.80 \pm   0.37$ & $ 30.06 \pm   0.73$ & $  0.37 \pm   0.22$ & 299.52/346 & $   8.05 \pm   0.13$ &  \nodata & 325.72/347 \\
416.65 & 416.67 & $ 13.31 \pm   0.37$ & $ 31.18 \pm   0.71$ & $  0.37 \pm   0.23$ & 275.05/346 & $   8.20 \pm   0.14$ &  \nodata & 291.28/347 \\
416.67 & 416.69 & $ 13.19 \pm   0.35$ & $ 30.39 \pm   0.74$ & $  0.37 \pm   0.21$ & 296.96/346 & $   8.03 \pm   0.13$ &  \nodata & 322.23/347 \\
416.69 & 416.72 & $ 11.49 \pm   0.32$ & $ 32.35 \pm   0.69$ & $  0.37 \pm   0.23$ & 258.03/346 & $   8.49 \pm   0.15$ &  \nodata & 268.30/347 \\
416.72 & 416.76 & $  7.37 \pm   0.21$ & $ 29.59 \pm   0.70$ & $  0.37 \pm   0.24$ & 341.34/346 & $   7.77 \pm   0.14$ &  \nodata & 354.76/347 \\
416.76 & 416.80 & $  7.33 \pm   0.20$ & $ 30.49 \pm   0.75$ & $  0.37 \pm   0.22$ & 322.79/346 & $   8.02 \pm   0.14$ &  \nodata & 345.60/347 \\
416.80 & 416.84 & $  7.08 \pm   0.20$ & $ 30.58 \pm   0.74$ & $  0.37 \pm   0.23$ & 347.43/346 & $   8.06 \pm   0.14$ &  \nodata & 364.61/347 \\
416.84 & 416.89 & $  6.49 \pm   0.18$ & $ 29.40 \pm   0.69$ & $  0.37 \pm   0.23$ & 288.42/346 & $   7.74 \pm   0.14$ &  \nodata & 302.85/347 \\
416.89 & 416.94 & $  5.82 \pm   0.16$ & $ 29.23 \pm   0.73$ & $  0.37 \pm   0.22$ & 375.91/346 & $   7.66 \pm   0.14$ &  \nodata & 394.02/347 \\
416.94 & 417.00 & $  4.91 \pm   0.14$ & $ 27.97 \pm   0.73$ & $  0.37 \pm   0.24$ & 399.67/346 & $   7.39 \pm   0.14$ &  \nodata & 418.61/347 \\
417.00 & 417.07 & $  4.37 \pm   0.12$ & $ 27.05 \pm   0.75$ & $  0.37 \pm   0.24$ & 390.55/346 & $   7.15 \pm   0.14$ &  \nodata & 408.53/347 \\
417.07 & 417.16 & $  3.42 \pm   0.10$ & $ 26.69 \pm   0.73$ & $  0.37 \pm   0.24$ & 339.64/346 & $   7.08 \pm   0.14$ &  \nodata & 356.46/347 \\
417.16 & 417.26 & $  3.27 \pm   0.09$ & $ 26.50 \pm   0.78$ & $  0.37 \pm   0.23$ & 384.28/346 & $   7.09 \pm   0.13$ &  \nodata & 411.47/347 \\
417.26 & 417.37 & $  2.68 \pm   0.08$ & $ 23.73 \pm   0.74$ & $  0.37 \pm   0.26$ & 309.26/346 & $   6.37 \pm   0.13$ &  \nodata & 330.30/347 \\
417.37 & 418.17 & $  0.47 \pm   0.02$ & $ 23.69 \pm   1.29$ & $  0.37 \pm   0.28$ & 437.20/346 & $   6.49 \pm   0.18$ &  \nodata & 473.36/347 \\
418.17 & 420.76 & $  0.16 \pm   0.01$ & $ 27.46 \pm   2.58$ & $  0.37 \pm   0.34$ & 369.33/346 & $   3.97 \pm   0.29$ & $ 14.96 \pm   1.54$ & 348.84/345 \\
420.76 & 425.32 & $  0.09 \pm   0.01$ & $ 17.29 \pm   6.83$ & $  0.37 \pm   0.39$ & 433.69/346 & $   5.76 \pm   0.35$ &  \nodata & 452.90/347 \\
425.32 & 425.36 & $  8.10 \pm   0.22$ & $ 29.31 \pm   0.74$ & $  0.37 \pm   0.22$ & 306.44/346 & $   7.80 \pm   0.13$ &  \nodata & 332.35/347 \\
425.36 & 431.63 & $  0.06 \pm   0.01$ & $ 20.81 \pm   8.03$ & $  0.37 \pm   0.46$ & 258.44/346 & $   5.30 \pm   0.41$ &  \nodata & 272.22/347 \\
431.62 & 431.74 & $  9.73 \pm   0.13$ & $ 32.38 \pm   0.42$ & $  0.37 \pm   0.10$ & 478.50/346 & $  6.12 \pm   0.25$ & $ 12.85 \pm   0.58$ & 451.62/345 \\
431.74 & 431.87 & $ 19.99 \pm   0.19$ & $ 35.92 \pm   0.29$ & $  0.37 \pm   0.07$ & 578.67/346 & $   6.77 \pm   0.24$ & $ 13.08 \pm   0.44$ & 556.25/345 \\
431.87 & 432.93 & $  0.43 \pm   0.02$ & $ 29.18 \pm   1.16$ & $  0.37 \pm   0.25$ & 385.81/346 & $   7.61 \pm   0.19$ &  \nodata & 422.69/347 \\
432.93 & 432.96 & $ 12.11 \pm   0.30$ & $ 30.70 \pm   0.61$ & $  0.37 \pm   0.22$ & 323.04/346 & $   8.05 \pm   0.13$ &  \nodata & 334.62/347 \\
432.96 & 433.02 & $  3.01 \pm   0.11$ & $ 26.76 \pm   0.93$ & $  0.37 \pm   0.32$ & 350.45/346 & $   7.08 \pm   0.17$ &  \nodata & 361.31/347 \\
\enddata